\documentclass[12pt,preprint2]{emulateapj}
\usepackage{graphicx}
\usepackage{url}
\usepackage{apjfonts}

\newcommand{\degree}{^{\circ}}
\newcommand{\e}{\varepsilon}

\newcommand{\rg}{\ensuremath{r_\mathrm{g}}}
\newcommand{\themis}{\ensuremath{\theta_\mathrm{e}}}

\shorttitle{Improved Reflection Models of Accretion Disks}
\shortauthors{Garc\'{\i}a \& et al.}

\begin{document}

\title{Improved Reflection Models of Black-Hole Accretion Disks: Treating the
Angular Distribution of X-rays}

\author{J. Garc\'ia\altaffilmark{1}, T. Dauser\altaffilmark{2},
  A. Lohfink\altaffilmark{3}, T.~R. Kallman\altaffilmark{4},
  J. Steiner\altaffilmark{1}, J.~E. McClintock\altaffilmark{1},
  L.~Brenneman\altaffilmark{1} J. Wilms\altaffilmark{2},
  W. Eikmann\altaffilmark{2}, C.~S. Reynolds\altaffilmark{3},
  F. Tombesi\altaffilmark{2}}

\altaffiltext{1}{Harvard-Smithsonian Center for Astrophysics, 60
  Garden St., Cambridge, MA 02138 USA; javier@head.cfa.harvard.edu,
  jem@cfa.harvard. edu, lbrenneman@cfa.harvard.edu, }

\altaffiltext{2}{Dr. Karl Remeis-Observatory and Erlangen Centre for
  Astroparticle \\ Physics, Sternwartstr. 7, 96049 Bamberg, Germany;
  thomas.dauser@ sternwarte.uni-erlangen.de,
  joern.wilms@sternwarte.uni-erlangen.de,
  wiebke.eikmann@sternwarte.uni-erlangen.de}

\altaffiltext{3}{Department of Astronomy, University of Maryland,
  College Park, MD, USA \& Joint Space-Science Institute, University
  of Maryland, College Park, MD, ~USA; alohfink@astro.umd.edu,
  chris@astro.umd.edu, ftombesi@astro.umd.edu}

\altaffiltext{4}{NASA Goddard Space Flight Center, Greenbelt, MD
  20771 USA; timothy.r.kallman@nasa.gov}

%==================================================================================
%
\begin{abstract}

  X-ray reflection models are used to constrain the properties of the
  accretion disk, such as the degree of ionization of the gas and the
  elemental abundances. In combination with general relativistic ray
  tracing codes, additional parameters like the spin of the black hole and
  the inclination to the system can be determined. However, current
  reflection models used for such studies only provide angle-averaged
  solutions for the flux reflected at the surface of the disk. Moreover, 
  the emission angle of the photons changes over the disk due to relativistic 
  light bending. To overcome this simplification, we have constructed 
  an angle-dependent reflection model with the {\sc xillver} code and
  self-consistently connected it with the relativistic blurring code 
  {\sc relline}. The new model, {\tt relxill}, calculates 
  the proper emission angle of the radiation at each point on￼ the accretion 
  disk, and then takes the corresponding reflection spectrum into account.
  We show that the reflected spectra from illuminated disks follow a 
  limb-brightening law highly dependent on the ionization of disk and 
  yet different from the commonly assumed form $I \propto \ln(1+1/\mu)$.
  A detailed comparison with the angle-averaged model is carried out in 
  order to determine the bias in the parameters obtained by fitting a 
  typical relativistic reflection spectrum. These simulations reveal that 
  although the spin and inclination are mildly affected, the Fe abundance 
  can be over-estimated by up to a factor of two when derived from
  angle-averaged models. The fit of the new model to the {\it Suzaku} 
  observation of the Seyfert galaxy Ark~120 clearly shows a significant
  improvement in the constrain of the physical parameters, in particular 
  by enhancing the accuracy in the inclination angle and the spin
  determinations.

\end{abstract}
%
%==================================================================================
%
\section{Introduction}
Despite the huge difference in mass, Galactic black holes (GBH) in binary
systems and super-massive black holes in active galactic nuclei (AGN) show
similar X-ray properties. In particular, in many sources an emission component
is present, which is generated by high-energy coronal photons that are
reprocessed in an optically-thick accretion disk. The most prominent feature in
this component -- commonly referred to as the ``reflection'' spectrum -- is the
Fe K emission line at 6--7~keV, which is produced by fluorescence. This
reflection spectrum from the inner disk region carries information on the
physical composition and condition of the matter in strong gravitational fields
\citep[e.g.,][]{fab00,fab03,rey03,dov04,mil08}. Doppler effects, light bending
and gravitational redshift skew the Fe line (and the fluorescence lines of
other elements) and can, for rapidly spinning black holes, extend the red wing
of the line to very low energies \citep{fab82,lao91,dab97,dov04,bre06,dau10}.
Hence, in order to fit observations, the pure reflection spectrum has to be
convolved by relativistic blurring algorithms that account for these
relativistic effects \citep[e.g.,][]{mil08,ste11,rey12,fab12b,Dauser2012a}. This
relativistic smearing highly depends on the parameters of the system, like the
black hole spin and the inclination of the system.
However, most current reflection models used for such studies only
provide an angle-averaged solution for the flux reflected at the surface of the
disk, which can systematically affect the inferred disk parameters.
Furthermore, an isotropic distribution or a particular limb-darkening law is
often chosen \citep[e.g.][]{chi12,wal13}, despite the fact that radiative
transfer calculations predict an emission excess near grazing angles
\citep[known as limb brightening, e.g.,][]{roz08,svo09,roz11}. In fact, the
pioneer and widely used relativistic line convolution model {\tt laor}
\citep{lao91} intrinsically assumes a limb-darkening law.

Several authors have studied relativistic effects on the emission of
X-rays from accretion disks, while considering the angular
distribution of the radiation. \cite{mar00} performed calculations of
the relativistic effects acting on both the reflection continuum and
the iron line in a Kerr-metric, and they showed that for rapidly
spinning black holes the line equivalent width is substantially
enhanced. \cite{rey00} discussed (within the context the analysis of
{\it ASCA} data for NGC~4258) how gravitational light bending can
affect the Fe line emission in high-inclination sources. \cite{bec04}
and \cite{dov04} presented similar calculations emphasizing how the
assumed angular emissivity law (limb darkening or brightening) can
significantly affect the line profile when relativistic effects are
important. These studies were extended by \cite{nie08} in order to
explain the Fe line profile and variability observed in
MCG$-$6-30-15. More recently, \cite{svo09} performed relativistic
radiative transfer calculations of X-ray irradiated disk atmospheres
with the {\sc noar} code to determine the impact of light bending on
X-rays in the 2--10~keV band. They concluded that the uncertainty in
the angular distribution of reflected radiation can translate into a
$\sim 20\%$ uncertainty in the determination of $R_\mathrm{in}$, and
thus also in the spin parameter $a_*$. The emphasis of their work is
on the relativistic effects and the estimation of uncertainties, and
they omit a detailed ionization-balance calculation of the reflection
spectrum.

In the past, X-ray reflection from a cold, neutral slab for the K lines of
heavy elements has been discussed extensively in the context of X-ray binaries
\citep{bas78} and AGN \citep{geo91,mat91}. Additional earlier work treats the
angular-dependence of electron scattering in cold material \citep{ghi94}, and
in hot thermal plasmas \citep{haa93}, while neglecting photoelectric absorption
and line emission. \cite{mat96} carried out detailed Monte Carlo calculations
of the Fe K$\alpha$ emission from X-ray photoionized accretion disks, showing
that the line strength and shape depend on both the ionization stage of the
material and on the disk inclination angle. \cite{vrt93} computed the disk
corona structure in X-ray binaries under the assumptions of ionization,
thermal, and hydrostatic balance of gas illuminated by the central continuum
source. They explained the Fe K emission and the absorption edge in terms of
the inclination angle and the shape of the incident source spectrum. More
sophisticated ionized reflection calculations have been presented by
\cite{roz08} in the context of accretion disks around supermassive black holes;
their models depict the effects of a limb-brightening law on the reflected
X-rays. The reflection model that has been most widely used by observers, for
both general application and for measuring black hole spin via the Fe-line
method, is {\sc reflionx} \citep{ros05}. This industry-standard model solves
the radiation transfer problem using a diffusion equation, which imposes the
limitation that the model can only deliver an angle-averaged spectrum.  

We overcome the simplifications adopted by previous models -- i.e., limited
ionization balance calculations and angle-averaged reflected flux -- by
exploiting the full capabilities of our reflection code {\sc xillver}
\citep{gar10, gar13a}, to treat the angular distribution of the reflected
X-rays. The Feautrier method we employ \citep{fea64} enables {\sc xillver} to
calculate the specific intensity of the radiation field as a function of
energy, position in the slab, and viewing angle. This allows us to construct a
grid of reflection models in which the inclination angle is included as an
explicit fitting parameter. Furthermore, using this approach we are able to
address a further complication. Due to general relativistic light-bending,
photons are emitted over a wide range of angles depending on their location on
the accretion disk. This complication can be solved by directly linking the
reflection code to a relativistic smearing kernel. Accordingly, we use the
angle-dependent {\sc xillver} reflection code to assign a proper reflected
spectrum for each point on the disk, which then depends on the actual emission
angle derived from general relativity. These single spectra are then
relativistically smeared using {\sc relline} \citep{dau10,dau13} before being
integrated to yield the total reflection spectrum. As each of these emission
points has to be treated separately, a simple combination of {\sc xillver} and
{\sc relline} will yield inconsistent results.

In this paper we present for the first time a complete description of the
spectrum reflected by an ionized accretion disk around a black hole that is
complete, self-consistent, and takes into account the angular distribution of
the reflected X-rays. The new model, {\tt relxill}, is provided in the
standard format, enabling its use with commonly employed fitting packages such
as {\sc xspec} \citep{arn96} and {\sc isis} \citep{hou00}; the model's input
parameters are the same as that of its parent model {\tt relconv}. An
additional version of the new model, {\tt relxill\_lp}, which simulates
the reflected spectra assuming a lamppost geometry, is also supplied.

This paper is organized as follows: In Section~\ref{sec:xill}, we describe our
angle-dependent solution for the reflected spectrum and compare it with the
commonly used angle-averaged solution.  Sections~\ref{sec:acc} and
\ref{sec:ang} first explore how relativistic effects modify the reflected
spectrum, and they then discuss how we integrated the reflection code {\sc
xillver} with the relativistic blurring code {\sc relline}. In
Section~\ref{sec:sim}, we assess, via a detailed error analysis, how the
angle-averaged solution yields biased values of the various model parameters.
An application of our model to an analysis of {\it Suzaku} data for Ark~120 is
discussed in Section~\ref{sec:obs}, and we offer our conclusions in
Section~\ref{sec:con}.
%
%==================================================================================
%
\section{Angular Solution of the Reflected Spectrum}\label{sec:xill}
The X-ray reflection calculations presented in this paper are based upon
those from \cite{gar13a}, with the difference being that we now exploit the full
capabilities of our reflection code {\sc xillver} by extracting the
solution of the emergent spectra for individual viewing angles. These models 
and the details of the numerical code have been extensively described
by \cite{gar10}, \cite{gar11}, and \cite{gar13a}. Here we will discuss
the theoretical aspects concerning the angular dependence of the
radiation fields, and their effects on the observed spectrum.

\begin{figure}
\epsscale{1.0}\plotone{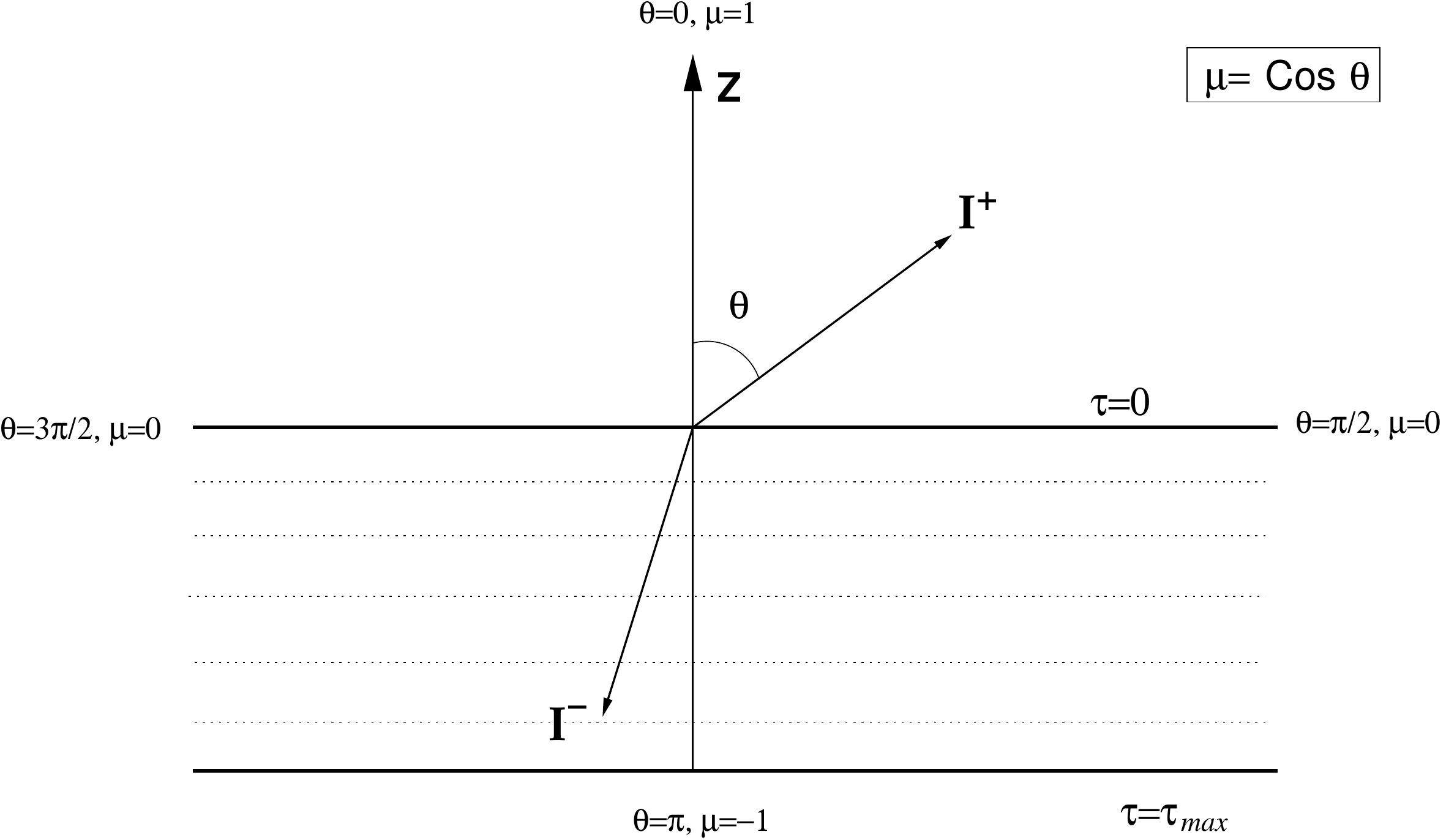}
\caption{Plane-parallel slab with the incoming and outgoing radiation fields.
}
\label{cartoon}
\end{figure}

The basic layout of the reflection problem is shown in the diagram of
Figure~\ref{cartoon}. The gas in the atmosphere of the accretion disk 
(i.e., within a few Thomson depths) is assumed to be at a density of
$n_{\mathrm H}=10^{15}$~cm$^{-3}$, in a plane-parallel geometry, 
where the coordinate of importance is shown in the vertical 
direction perpendicular to the slab ($z$-axis).
The surface of the disk is set at $\tau_{\mathrm T}=0$,
and the base of the atmosphere at $\tau_\mathrm{max}=10$, where 
$d\tau_{\mathrm T}=1.2n_{\mathrm H} \sigma_{\mathrm T}dz$ is the Thomson
optical depth, and $\sigma_{\mathrm T}=6.65\times 10^{-25}$~cm$^2$ is the
Thomson classical cross section for electron scattering.

The disk is illuminated by a primary source of X-ray radiation from
above. The radiation enters the slab at $\tau_{\mathrm T}=0$, and is
reprocessed by the atmosphere. The intensity of the radiation field $I(z,\mu,\e)$
at each position $z$ the atmosphere, per energy $\e$, unit area, and steradian
is determined by solving the time-independent, one-dimensional plane-parallel
radiation transfer equation
\begin{equation}\label{ertz}
\mu \frac{\partial I_{\e}(z,\mu)}{\partial z} = \eta_{\e}(z) - \chi_{\e}(z) I_{\e}(z,\mu)
\end{equation}
where $\mu$ is the cosine of the angle $\theta$ with respect to the disk normal,
and $\eta_{\e}(z)$ and $\chi_{\e}(z)$ are the total emissivity and opacity, respectively.

At each point in the slab one can distinguish between the incoming and the
outgoing components of the radiation for $\pm \mu$ (see Figure~\ref{cartoon}),
and thus write two separate transfer equations
\begin{equation}\label{ertp}
\mu\frac{\partial I^+_{\e}(\tau_{\e},\mu)}{\partial \tau_{\e}} = I^+_{\e}(\tau_{\e},\mu) - S_{\e}(\tau_{\e})
\end{equation}
\begin{equation}\label{ertm}
- \mu\frac{\partial I^-_{\e}(\tau_{\e},\mu)}{\partial \tau_{\e}} = I^-_{\e}(\tau_{\e},\mu) - S_{\e}(\tau_{\e})
\end{equation}
where $I^+_{\e}$ and $I^-_{\e}$ refer to positive and negative values of
$\mu$, respectively, and $\tau_{\e} = -\chi_{\e}(z)dz$ is the total optical depth.
By defining the symmetric and anti-symmetric averages
\begin{equation}\label{eu}
u_{\e}(\tau_{\e},\mu) = \frac{1}{2}\left[ I^+_{\e}(\tau_{\e},\mu) + I^-_{\e}(\tau_{\e},\mu) \right]
\end{equation}
\begin{equation}\label{ev}
v_{\e}(\tau_{\e},\mu) = \frac{1}{2}\left[ I^+_{\e}(\tau_{\e},\mu) - I^-_{\e}(\tau_{\e},\mu) \right]
\end{equation}
the transfer equation (\ref{ertz}) can be rewritten as
\begin{equation}\label{ertu}
\mu^2 \frac{\partial^2 u_{\e}(\tau_{\e},\mu)}{\partial \tau_{\e}^2} = u_{\e}(\tau_{\e},\mu) - S_{\e}(\tau_{\e}),
\end{equation}
where $S_{\e}(\tau_{\e}) = \eta_{\e}/\chi_{\e}$ is the source function. The line plus continuum
emissivities and opacities depend on the structure of the gas, which is
determined at each point in the slab by the photoionization code {\sc xstar}
\citep{kal01}. Details on the calculation of these quantities and on the
atomic data implemented in these calculations are fully described in \cite{gar10}
and \cite{gar13a}.
Expression~(\ref{ertu}) is a second-order differential equation subject to
two boundary conditions. At the top ($\tau_{\e}=0$), the incoming
radiation field $I^-_{\e}(0,\mu)$ is known. Since
$u_{\e} - v_{\e} = I^-_{\e}$, and 
$v_{\e} = \mu \frac{\partial u_{\e}}{\partial \tau_{\e}}$, the boundary
condition at the surface of the slab can be expressed as
\begin{equation}\label{ebc1}
u_{\e}(0,\mu) - \mu \left( \frac{\partial u_{\e}}{\partial \tau_{\e}}\right)_{0} = I_\mathrm{inc}.
\end{equation}
Assuming no irradiation from below, we set  $I^+_{\e}(\tau_\mathrm{max},\mu)=0$ 
at the bottom of the atmosphere ($\tau_{\e}=\tau_\mathrm{max}$).
Thus, using $u_{\e} + v_{\e} = I^+_{\e}$, the boundary
condition at the bottom of the slab becomes
\begin{equation}\label{ebc2}
u_{\e}(\tau_\mathrm{max},\mu) + \mu \left( \frac{\partial u_{\e}}{\partial \tau_{\e}}\right)_{\tau_\mathrm{max}} = 0
\end{equation}
and $\tau_\mathrm{max} = 10 \tau_\mathrm{T}$ is chosen for all the calculations.

The transfer equation (\ref{ertu}) and its boundary conditions (\ref{ebc1}) 
and (\ref{ebc2}) are converted into a set of difference equations via the 
discretization of all variables, in the form
\begin{equation}\label{erti}
\mu^2 u_i'' = u_i - S_i
\end{equation}
where $i=1,\ldots,N$ denotes a particular position in a grid of $N$
points $\tau_i$. Notice that, for simplicity, we have intentionally suppressed
the explicit dependence on $\e$ and $\mu$.

The derivatives of $u_i$ in a non-uniform mesh can 
be found using the Taylor's expansions
\begin{equation}\label{efor}
u_{i+1} = u_i + u_i' \Delta\tau_i + u_i'' \frac{\Delta\tau_i^2}{2} + ...
\end{equation}
\begin{equation}\label{ebac}
u_{i-1} = u_i - u_i' \Delta\tau_{i-1} + u_i'' \frac{\Delta\tau_{i-1}^2}{2} + ...
\end{equation}
which are usually referred to as the forward and backward finite differences.
Solving for $u_i'$ in (\ref{efor}) and substituting in (\ref{ebac}), the second
derivative can be expressed as
\begin{equation}\label{esed}
u_i'' = \frac{2}{\Delta\tau_i + \Delta\tau_{i-1}} \left[
u_{i-1} \left(\frac{1}{\Delta\tau_{i-1}} \right)
- u_i \left(\frac{\Delta\tau_i + \Delta\tau_{i-1}}{\Delta\tau_i\Delta\tau_{i-1}}  \right)
u_{i+1} \left(\frac{1}{\Delta\tau_i}\right) 
\right].
\end{equation}

This can now be inserted into (\ref{erti}) to write
\begin{eqnarray}
u_{i-1}\left[\frac{-2\mu^2}{\Delta\tau_{i-1}(\Delta\tau_i + \Delta\tau_{i-1})} \right]
 &+& u_{i}\left[\frac{2\mu^2}{\Delta\tau_{i}\Delta\tau_{i-1}} \right] \\
 &+& u_{i+1}\left[\frac{-2\mu^2}{\Delta\tau_{i}(\Delta\tau_i + \Delta\tau_{i-1})} \right]
 = S_i \nonumber
\end{eqnarray}

Equation~(\ref{esed}) can be inserted into (\ref{efor}) to get a similar
expression for the first derivative in terms of $u_{i-1}, u_i,$ and $u_{i+1}$.
However, for the boundary condition at the top
\begin{equation}\label{ebc}
u_1 - \mu u_1' = I_\mathrm{inc}
\end{equation}
a different expression is needed. An alternative
is to rely upon the second derivative provided by 
equation (\ref{erti}). Thus
\begin{equation}
u_i' = u_{i+1}\left(\frac{1}{\Delta\tau_i}\right) 
- u_i \left(\frac{1}{\Delta\tau_i} + \frac{\Delta\tau_i}{2\mu^2} \right)
       + S_i \frac{\Delta\tau_i}{2\mu^2}
\end{equation}
and (\ref{ebc}) becomes
\begin{equation}
u_1 \left( 1 + \frac{\mu}{\Delta\tau_1} + \frac{\Delta\tau_1}{2\mu} \right)
- u_2 \left(\frac{\mu}{\Delta\tau_1}\right)
= I_\mathrm{inc} + S_1 \frac{\Delta\tau_1}{2\mu}
\end{equation}
A similar expression can be found for the inner boundary at
$\tau=\tau_\mathrm{max}$.

The numerical solution in this approach proceeds by a forward elimination plus
backward substitution scheme. We start using the boundary condition at the
surface to express $u_1$ in terms of $u_2$. The result is then applied to
variables $u_1$, $u_2$ and $u_3$ in order to express $u_2$ in terms of $u_3$.
Iteratively, and in the same way, we can express each $u_i$ in terms of
$u_{i+1}$. We finally reach the last point in the grid, and the boundary
condition for $\tau_\mathrm{max}$ gives $u_N$. Back substitutions then produce
$u_N \rightarrow u_{N-1} \rightarrow u_{N-2} \rightarrow \cdots \rightarrow u_2
\rightarrow u_1$, so all $u_i$ are known.

Once the solution for all $u_i$ is known, it is trivial to obtain the outgoing
intensity $I^+(0)$ at the surface of the slab (the reflected component), by
simply using (\ref{eu})
\begin{equation}
I^+_1 = 2u_1 - I^-_1
\end{equation}
where, again, $I^-_1 = I_\mathrm{inc}$. This is the solution of the reflected
intensity at the illuminated surface of the slab for every energy $E$ and every
angle $\mu$. For the calculations presented here, the grid in $\mu$ angles is
defined as
\begin{equation}
\mu_j = \frac{j-0.5}{M} \ \ \ \ \ \ \ \ \ (j = 1, 2,\ldots, M)
\end{equation}
with $M=10$, and $\Delta\mu = 1/M$. Note that for the {\sc xillver} models 
previously published \citep{gar10,gar11,gar13a}, the reflected spectra were
provided in terms of the {\it angle-averaged} quantity, 
\begin{equation}\label{eave}
F_{\e}^+(0) = \frac{1}{2} \int_0^1 I^+_{\e}(0,\mu) \mu d\mu
\end{equation}
defined as the first moment of the radiation field (although we are only 
taking the component of the field for positives $\mu$, the one that reaches 
the observer).

\begin{figure*}
\epsscale{1.0}\plotone{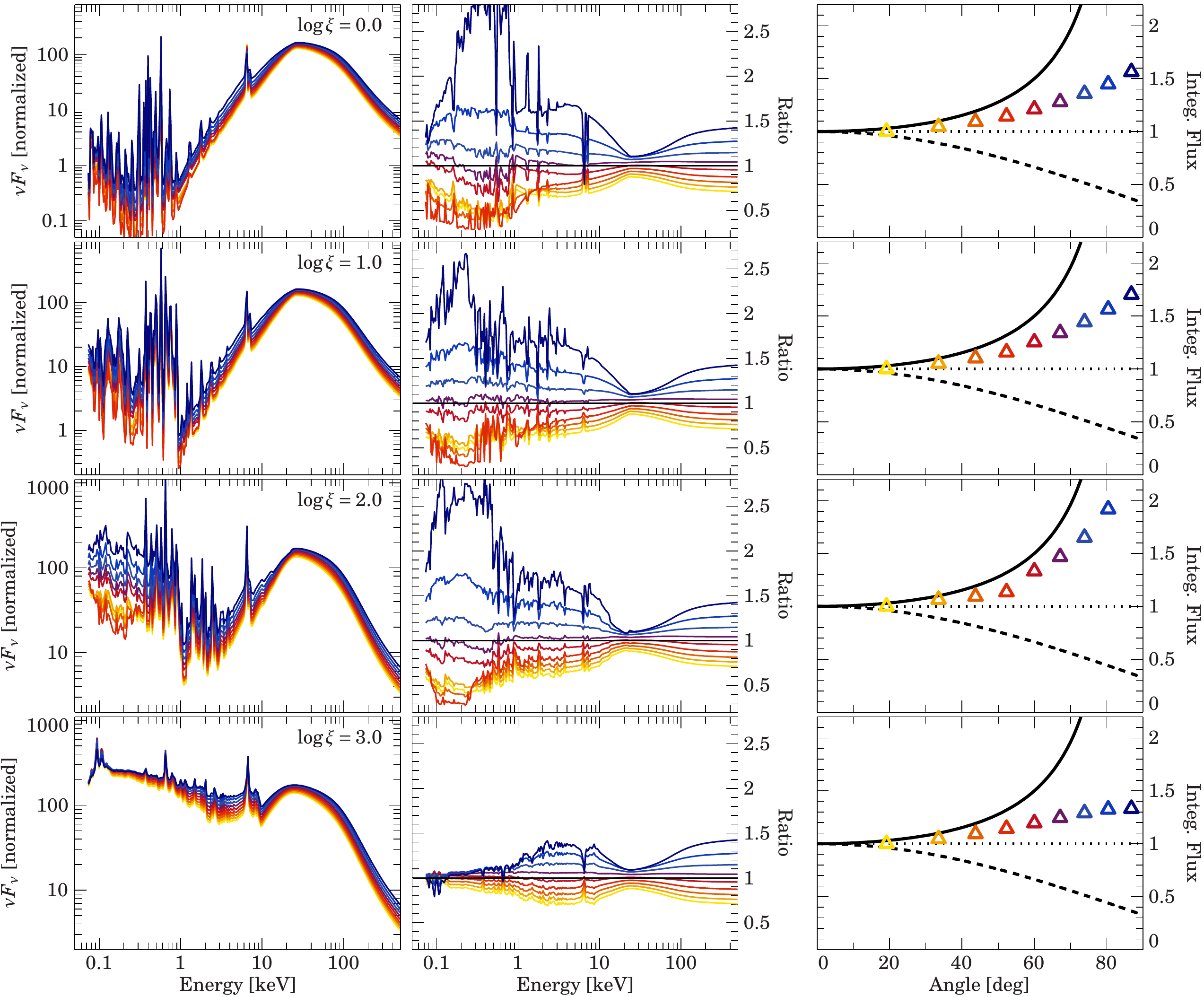}
\caption{Left panels: Angular distribution of the reflected spectrum
  from an accretion disk calculated with {\sc xillver}. Middle panels:
  ratio of the angular solution to the angle-average.  Right panels:
  Integrated flux as a function of the viewing angle. Each angular
  solution is identified with a different color. Each row of panels
  correspond to a particular ionization parameter, as indicated in the
  left panels. The solid and dashed lines represent the
  limb-brightening \citep[$I \propto \ln (1+1/\mu)$][]{haa93}
  and darkening laws \citep[$I \propto \ln (1+2.06\mu)$][]{lao91},
  respectively, which are commonly used in relativistic blurring
  kernels.}
\label{xillver}
\end{figure*}

The left column in Figure~\ref{xillver} shows the reflection spectra across all
viewing angles for different values of the ionization parameter, as indicated
in each panel. These are displayed in different colors (see right column for
an identification). At first glance the spectra look very similar to each
other. However, the ratio between spectra for individual $\mu$ and the
angle-averaged spectrum reveals differences that can reach up to a factor 2.5.
Especially interesting is that the spectral features behave differently than
the continuum. This makes sense because the viewing angle affects the intensity
of the emergent radiation by changing the {\it effective optical depth}, given
by the geometrical projection of the optical depth along the line of sight
according to 
\begin{equation}\label{eteff}
\tau_\mathrm{eff} = \tau/\mu \quad.
\end{equation} 
An observer looking at grazing angles (small $\mu$) will effectively see a
larger optical depth. Or in other words, photons emitted at a particular depth
in the slab will see a larger effective depth to escape out of the disk at
grazing angles. Therefore, looking at the energy distribution of the emergent
radiation, spectral features that are originally produced at large $\tau$ (such
as those from  Fe and Ni) will be more attenuated than those that are produced
near the surface (such as C, N, and O  lines). This is also the case for
photons in the continuum, which is the reason for the Compton hump above
$20$~keV being strongly affected by the viewing angle
\citep{lig80,lig81,lig88,mag95}. In fact, high energy photons are the most
affected, since they experience a larger number of scatterings (thus, they will
be very sensitive to any small change in the optical depth).

The angular dependence of the reflected spectra for energies above $\sim
20$~keV is the same for all ionization parameters, since Compton scattering is
the main source of opacity, and these energies are insensitive to the gas
temperature. Conversely, the impact of the viewing angle at lower energies is
more pronounced where the photoelectric opacity becomes important. Because
low-ionization models are dominated by photoelectric absorption, angular
effects are accordingly stronger the lower the ionization parameter. An
important effect is the change in the slope of the reflected spectrum, 
as the the angular dependence of the solution is different at different energies. 
In general, models with an inclination of
$60\degree$-$70\degree$ seem to be the closest to the angle-averaged solution.
Therefore, we expect the models presented here to be most useful for systems
with either very low or very high inclination angles.

The energy dependence of the angular effects on the reflected spectrum becomes
very relevant when combined with relativistic blurring models, such as {\sc
relline}. Commonly, in these models one can choose between a limb-brightening
or a limb-darkening law which behaves uniformly for all energies
\citep{svo09,dau10,dau13}. In the right column of Figure~\ref{xillver}, we
compare the intensity predicted by our angular dependent simulation with these
empirical models. It is clear that our simulations tend towards limb
brightening, i.e., the integrated flux increases when the inclination angle
increases. For low ionization models, the predicted trend departs significantly
from the usual limb-brightening law, showing a much flatter profile. As the
ionization grows, the integrated flux predicted by {\sc xillver} becomes
steeper and approaches the limb-brightening law. However, for log~$\xi \gtrsim
3$, the profile becomes flatter, resembling isotropic behavior. This is due to
the fact that for high-ionization models, the ionization structure is dominated
by electron scattering, rather than photoionization and recombination. In this
regime, photons suffer a large number of scatterings in a hot atmosphere before
escaping towards the observer. The gas remains close to the Compton temperature
for a large range of optical depths \citep{gar13a}, meaning that an observer
will see a nearly isothermal atmosphere at any inclination. 
These results are of great relevance when contrasted with earlier
relativistic convolution models (e.g., {\tt kdblur, kerrconv}), 
which arbitrarily assume a limb-darkening law \citep{lao91}. 
Instead, all our models predict a limb-brightening behavior highly dependent 
on the particular ionization state of the gas.

\begin{figure}
\epsscale{1.0}\plotone{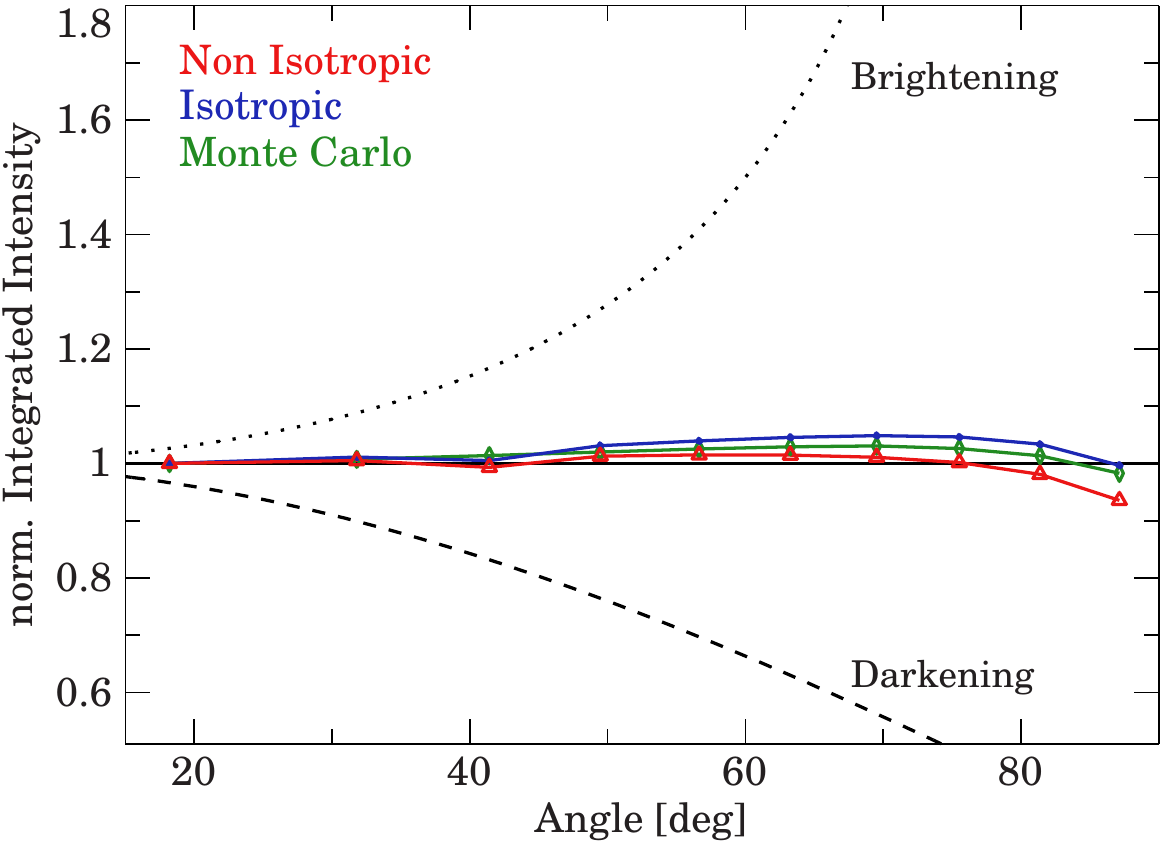}
\caption{Angular-dependence of the integrated flux emergent from an
  irradiated accretion disk. The limb-brightening and darkening laws
  are shown in green and gray for reference (see
  Figure~\ref{xillver}). A pure scattering {\sc xillver} calculation
  is shown in blue (circles), which follows an isotropic profile. A
  similar calculation (in red, triangles) that includes a phase
  function in the radiative transfer calculation shows a very similar
  profile.  The flatness is understood to reflect the intrinsic angle
  averaging which occurs due to many scatterings (i.e., the original
  emission angle information is lost). This result is confirmed with a
  Monte Carlo simulation under the same conditions, shown in green
  (diamonds).  }
\label{scattering}
\end{figure}

We have also explored the possibility of including an angular dependence in the
scattering term of the source function of Equation~(\ref{erti}). In the absence
of absorption and intrinsic emission (i.e., pure scattering), the source
function is simply reduced to $S_{\e}(\tau_{\e})=J_{\e}(\tau_{\e})$, where
$J_{\e}(\tau_{\e})=\int u_{\e}(\mu,\tau_{\e})d\mu$ is the zeroth moment of the
radiation field (i.e., the mean intensity), which is an angle-averaged
quantity. One can, however, introduce a phase function to describe the angular
properties of the electron scattering. A simple approximation in the
non-relativistic limit is the Rayleigh scattering phase function, for which the
source function can be expressed as
\begin{equation}
S_{\e}(\tau_{\e},\mu) = \frac{3}{8} [(3-\mu^2) J_{\e}(\tau_{\e}) + (3\mu^2-1) K_{\e}(\tau_{\e})]
\end{equation}
\citep[see Eq. 103, Chap. 1 of][]{cha60}, where $K_{\e}(\tau_{\e}) = \int
u_{\e}(\mu,\tau_{\e})\mu^2d\mu$ is the second moment of the radiation
field. Using this form of the source function we have calculated the
radiation transfer on an isothermal, pure scattering atmosphere.
Figure~\ref{scattering} shows that the angular distribution of the
integrated flux behaves very closely to the isotropic case.
This result demonstrates that the treatment of the Compton scattering
in {\sc xillver} is adequate for these kind of calculations and it is upheld by
3-dimensional Monte Carlo simulations which have shown that the reflection
intensity of a Compton scattering dominated atmosphere appears nearly isotropic
for all viewing angles.

%
%==================================================================================
%
\section{Accretion Disks in General Relativity}\label{sec:acc}
In this section we discuss the effects of strong gravity on photons
emitted from the surface of an accretion disk around compact objects,
and the numerical methods employed to predict the emitted spectrum
perceived by a distant observer. In the following we briefly
summarize how we calculate the general relativistic radiative
transfer around a rotating black hole. Accordingly, the equations of
motion in the \citet{Kerr1963} metric are used to describe the photon
and particle trajectories \citep{Bardeen1972}. The actual transfer is
solved by using ``Cunningham Transfer Function''
\citep{Cunningham1975}, as implemented by \citet{Speith1995}. This
approach is used by the {\sc relline} code, which was employed to
obtain the results presented in the following.\footnote{See \cite{dau10,dau13}
for more details on the calculation and the code.}

In the following, we use the dimensionless spin parameter $a_*$, which
takes values from $a_*=1$ (maximally rotating) to $a_*=0$ (non-spinning)
and $a_*=-1$ (negatively rotating, i.e., the black hole and the
accretion disk are counter rotating). Moreover, all lengths are
given in units of the gravitational radius, which is defined as
$\rg = GM/c^2$.

\begin{figure*}
\epsscale{1.0}\plotone{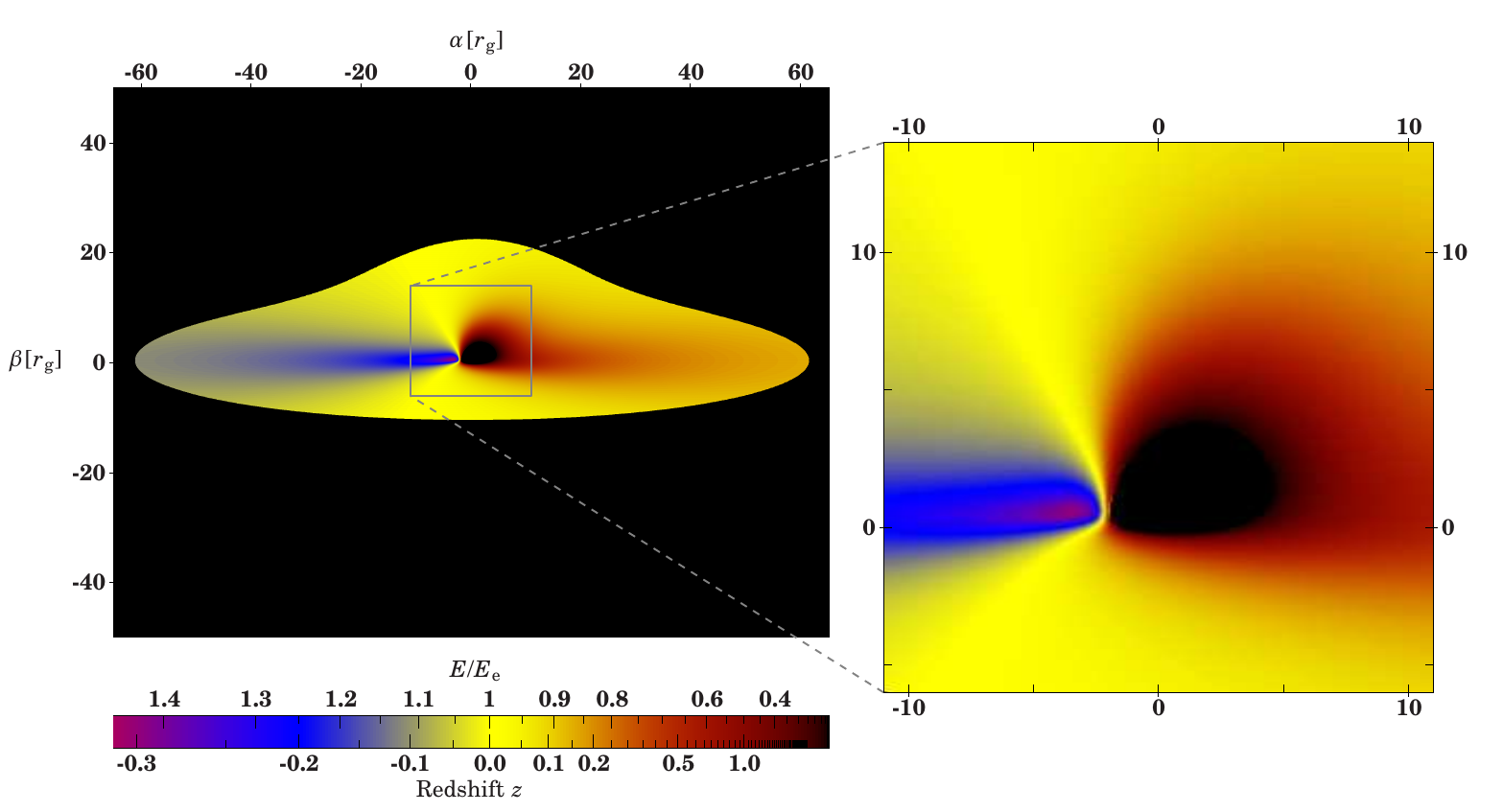}
\caption{Map of an accretion disk around a maximally spinning black
  hole $(a = 0.998)$ as seen by a distant observer at an inclination
  angle of $\theta = 80^\circ$. The disk ranges from the marginally stable
  radius $(r_\mathrm{in} = 1.24\,\rg)$ to $r_\mathrm{out} = 60\,\rg$. $\alpha$
  and $\beta$ are the coordinates defined on the plane of the sky
  (i.e., perpendicular to the line of sight; see
  \citealp{Cunningham1973}, equation 28). The color scale shows the
  energy shift of the photons; the geometric asymmetries are due to relativistic
  light bending. The blue-shifted (left) part of the disk moves towards
  the observer, whereas the right part recedes from the observer.  }
\label{disk1}
\end{figure*}

Using the approach outlined above, we are able to calculate the apparent image
of the accretion disk on the sky a distant observer would see
\citep[e.g.][]{Cunningham1973}. Figure~\ref{disk1} shows how light-bending
creates apparent asymmetries in the shape and warps the disk towards the
observer. Moreover, the energy shift the photons experience from the disk to
the observer is illustrated as it varies over the disk \citep[see][for more
details and explicit formulae]{dau10}. The main effects which impact the energy
are the Doppler shift (which leads to the blue and red division of the disk)
and the gravitational redshift, which increases with proximity to the black
hole.

\begin{figure*}
\epsscale{1.0}\plotone{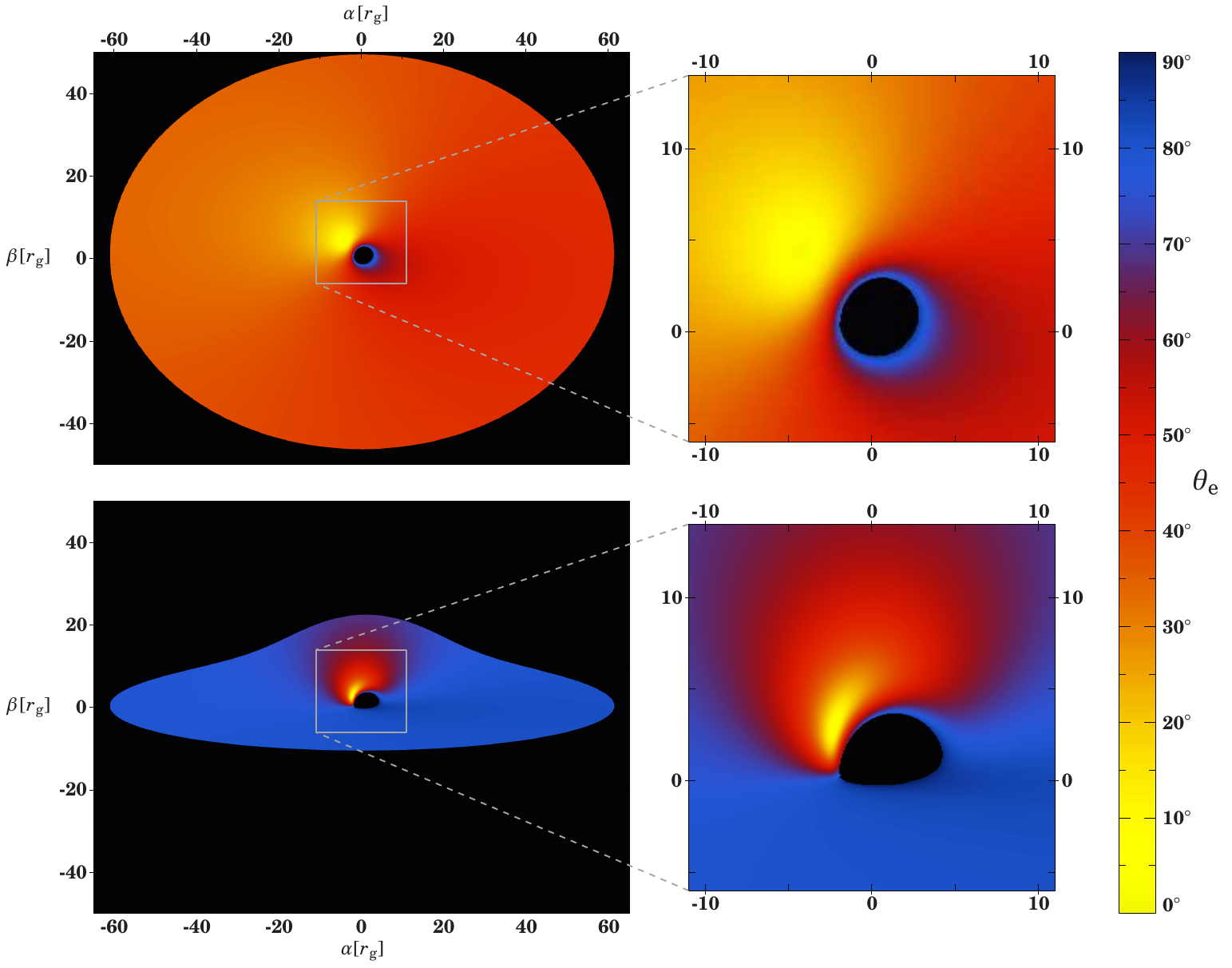}
\caption{Map of an accretion disk similar to Figure~\ref{disk1}, except
  that now the emission angle $\themis$ is depicted by the color scaling on
  the accretion disk. The disk is displayed for angles of $\theta=40\degree$
  (upper panel) and $\theta=80\degree$ (lower panel). While $\themis$
  takes almost any value close to the black hole, it is apparent that
  for increasing distance to the black hole this angle converges
  towards the inclination of the system.}
\label{disk2}
\end{figure*}
%==================================================================================

Besides the energy shift, general relativistic effects also alter the
direction of the photon, i.e., the photon is likely to be emitted at a
different angle than it is observed. From the momentum of the emitted
photon $\vec{p}_\mathrm{e}$ \citep[see][]{Bardeen1972}, the emission angle can be
easily calculated via
\begin{equation}\label{eq:emis_angle}
  \cos(\themis) =  \frac{\vec{p}_{\mathrm{e}\perp}}{|\vec{p}_\mathrm{e}|} \quad.
\end{equation} 
Figure~\ref{disk2} shows a map of $\themis$ on the surface of a disk
seen at inclinations of $\theta =40\degree$ and $\theta=80\degree$. In
both cases, the emission angle converges as expected towards the
viewing angle $\theta$ at sufficiently large distances from the
central black hole.  This is expected, as this implies that the photon
travels on a straight trajectory to the observer.  In particular for
$\theta=40\degree$, slight differences at larger distances are visible
between the left and the right sides, due to the rotation of the
accretion disk. Looking at the zoomed images in Figure~\ref{disk2}, it
is obvious that essentially every emission angle from $0\degree$ and $90\degree$ 
provides some contribution to the observed image. Specifically, $\themis$ is
significantly lower than the viewing angle for photons emitted behind
the black hole, as light bending effects are stronger in this case.
Even for $\theta=80\degree$, the emission angle approaches
$0\degree$ for a small part of the disk. Again, due to the rotation of
the disk and also the rotation of the black hole, this effect is
stronger for particles moving towards the observer and in the
rotational direction of the black hole. Therefore, the region of low
emission angles is in both cases shifted asymmetrically to the
left. Additionally, there is also a region of relatively high emission
angles at the innermost radii of the disk. This is best seen for
$\theta = 40\degree$, visible as a blue ring in Figure~\ref{disk2}. 
Note that although most of the emission angles in the figure are close to the
viewing angle, the inner parts radiate with much stronger intensity and
generally dominate the observation. In detail, for a standard
\citet{Shakura1973a} disk the emitted intensity behaves like $I \propto
r^{-3}$, while for a disk irradiated by a source on the black hole's rotational
axis, generally, an even steeper profile is expected
\citep[e.g.,][]{Fukumura2007a,Wilkins2012a,dau13}.
%
%==================================================================================
%
\section{Angle-Dependent Blurred Reflection}\label{sec:ang}
As discussed in Section~\ref{sec:xill}, the reflection spectra change
significantly depending on the inclination angle. Moreover, in the previous
section we showed that the emission angles are not necessarily equal (or even
close) to the inclination angle of the system. Therefore, it is imperative to
take into account the complete relativistic transfer function for each region
on the accretion disk, in order to properly predict the emitted spectrum. 

\subsection{General Properties (\texttt{relxill})} For this purpose, we
have extended the relativistic blurring code {\sc relline} and incorporated
angle-dependent reflection from the {\sc xillver} code. Accordingly, the model
behaves similarly to a convolution of the {\tt xillver} tables with
\texttt{relconv}. But instead of smearing the angle averaged reflection, this
new model, called \texttt{relxill}, combines the spectra from all points
in the image plane according to their observed emission angles, all while
properly smearing each point relativistically.  Although more physically
rigorous, under this new approach, no additional parameters enter the model.
In fact, because the angular solution of the reflected spectrum is
self-consistently calculated, no assumptions have to be made for the 
limb-brightening/darkening laws, and thus the {\tt limb} parameter is
removed from the model.
However, we have included a flag variable called {\tt angleon} to switch
between the new model with the full angular solution ({\tt angleon=1}), and the
old angle-average version ({\tt angleon=0}). This model is provided in
the appropriate format to be used in combination with other models included
in the commonly-used X-ray fitting packages such as {\sc xspec} \citep{arn96} 
and {\sc isis} \citep{hou00}.

\begin{figure}
\epsscale{1.0}\plotone{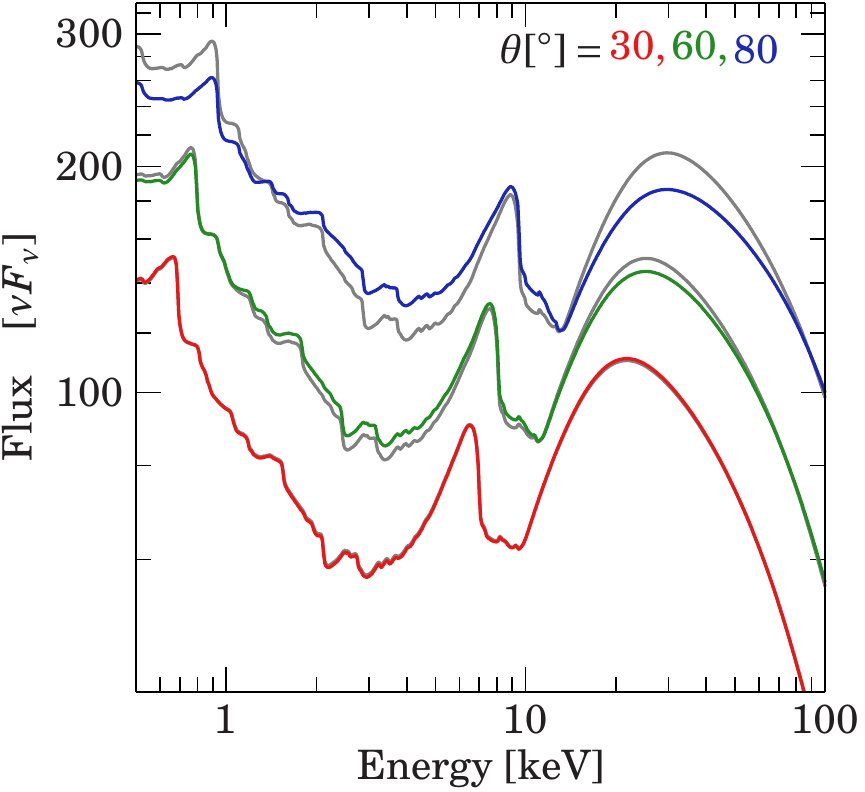}
\caption{Reflected spectrum calculated self-consistently with the new 
  model {\tt relxill} for different inclination angles. For each
  case, the prediction obtained with simple convolution of {\tt relconv}
  on the angle-dependent {\tt xillver} spectra (with angles linked) is 
  shown in grey, demonstrating the importance of the correct implementation
  of the angular solutions. Differences are bigger for larger inclinations
  because the relativistic effects are stronger. The other parameters are
  common to all spectra, i.e., ionization parameter log~$\xi=3$, photon 
  index $\Gamma=2$, emissivity $q=3$, and spin parameter $a_*=0.998$.
  }
\label{fig:comparison}
\end{figure}

It is important to clarify that a simple convolution of the angle-dependent
{\tt xillver} spectra with a relativistic blurring code will yield incorrect
results. As described in Section~\ref{sec:acc}, in presence of relativistic effects
photons emitted at many different angles can contribute to the spectrum 
observed at one particular inclination, due to strong light-bending. This 
effect is different at different locations on the surface of the accretion disk.
Therefore, the convolution of the reflected spectra needs to be done differently
at each particular radius, taking the correct contribution of the multiple 
angular solutions to the given viewing angle. Figure~\ref{fig:comparison} 
illustrates the differences in the final spectrum obtained from an simple
convolution of {\tt relconv} on the angle-dependent {\tt xillver} spectra with
the inclination angles linked together, and the one calculated
self-consistently with our new model {\tt relxill}.

\begin{figure*}
\epsscale{1.0}\plotone{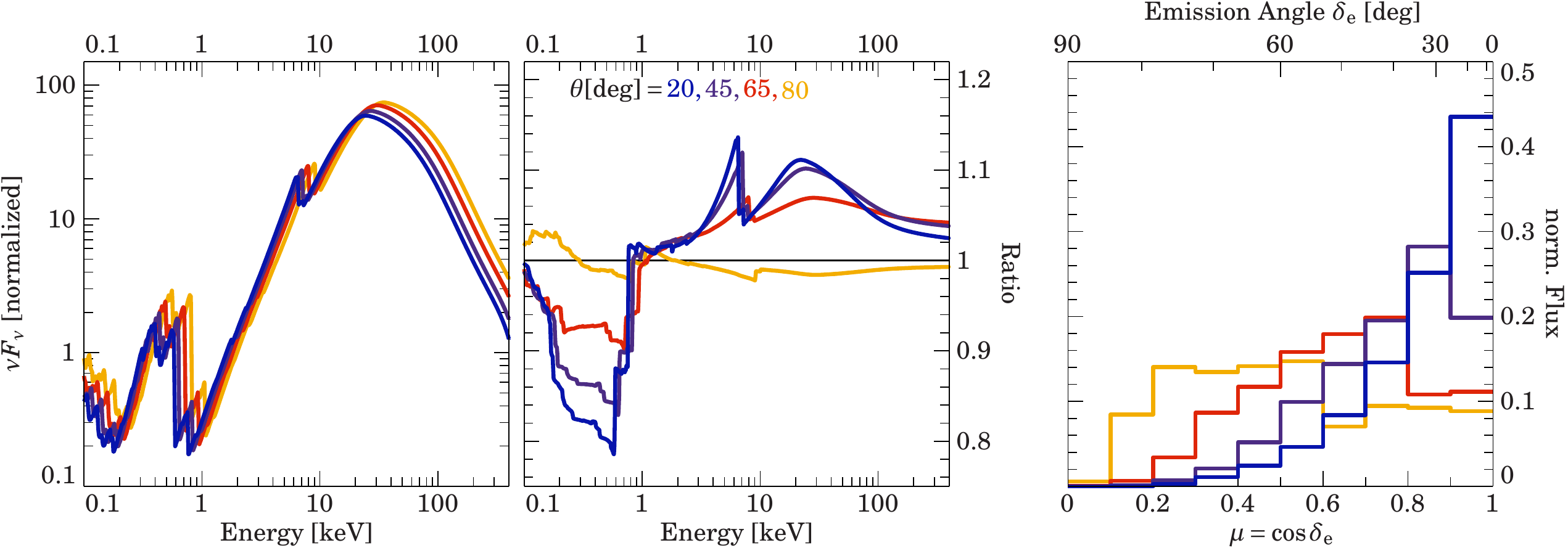}
\caption{Left panel: The \texttt{relxill} model for a typical parameter
  combination (ionization parameter $\xi=1$, photon index $\Gamma=2$,
  and the standard emissivity $I \propto r^{-3}$) for different
  inclinations. Middle panel: The ratio between \texttt{relxill} and the
  usual convolution of \texttt{xillver} with \texttt{relconv}. Right panel:
  Distribution of the flux across the range of emission angles. Note that a normal
  reflection code would assume a constant distribution here. }
\label{fig:relconv_xill}
\end{figure*}

In Figure~\ref{fig:relconv_xill} we show the results of the new model
\texttt{relxill} with typical parameter settings for different
inclinations of the system. As is well known and evident in the spectra in
the left panel, different inclinations primarily result in a net energy shift
for the spectrum.  However, when compared to the usual, averaged relativistic
convolution (middle panel) it is obvious that a deviation is present which is
highly dependent on the inclination. In general, these deviations are as large
as 20\% and distinct for lines and for the continuum. The reason for such
differences becomes evident in the right panel, which shows the distribution of
the flux per emission angle. In this particular case, there is a flat
distribution at large inclinations ($\theta=80\degree$), i.e., photons from
almost all angles are being observed with a similar contribution.
Consequently, the resulting spectra are also very similar. This is equivalent
to an angle-averaged solution, which is assumed by common reflection codes.  In
contrast, for small inclination angles the distribution is peaked at the actual
inclination angle. This leads to an intrinsically different reflection spectrum
and hence the relativistically smeared spectrum also differs significantly.
However, these results depend upon various parameter settings, like the
ionization of the disk ($\xi$), the abundances, the photon index $\Gamma$, the
emissivity $\epsilon$, or the spin $a_*$.

Note that while for low spin the relativistic effects are generally
less pronounced, this does not mean that the difference between the
average and the proper angle treatment vanishes. As it can be seen in
Figure~\ref{xillver}, even in the non-relativistic case, the averaged
spectrum coincides best with the spectrum emitted at
$\theta=65\degree$ and usually differs in shape for other smaller
and larger inclinations.

\subsection{Differences to the Angle-Average
  (\texttt{relxill\_lp})}

Recent measurements \citep{wilms2001a,fabian2002a,bre06,Dauser2012a,
fab12a,gal11,pon10,bre11,bre13,dur11,mil06,rei08} have revealed a steeper
emissivity at the inner parts of the accretion disk than the canonically
assumed $r^{-3}$ dependence \citep{Shakura1973a}. This is well understood in
the so-called ``lampost'' geometry \citep{mat91,Martocchia1996a}. For this
scenario, instead of coming from a corona around the inner regions of the disk
\citep{haa93,Dove1997a}, the primary radiation illuminating the disk comes from
a source placed on the rotational axis of the black hole \citep[e.g., ][for a
detailed discussion]{dau13}. Recent analyzes show that this model is capable of
explaining the observed spectra in some systems
\citep[e.g.,][]{Wilkins2011a,dur11,Dauser2012a}.

Due to the success of this model, we investigate the effect of a rigorous and
full treatment of the emission angle on the parameters describing the
reflection and the relativistic blurring. Under lampost geometry, the
emissivity profile is specified by the height $h$ of the primary source. While
a large height usually implies a flat emissivity in the inner region, it
steepens dramatically for decreasing height \citep[e.g., ][]{dau13}. This means
that for a source located near the black hole (low $h$), the inner portion of 
the accretion disk more strongly dominates the reflection spectrum than for 
a source located far away (high $h$).
 
\begin{figure*}
\epsscale{1.0}\plotone{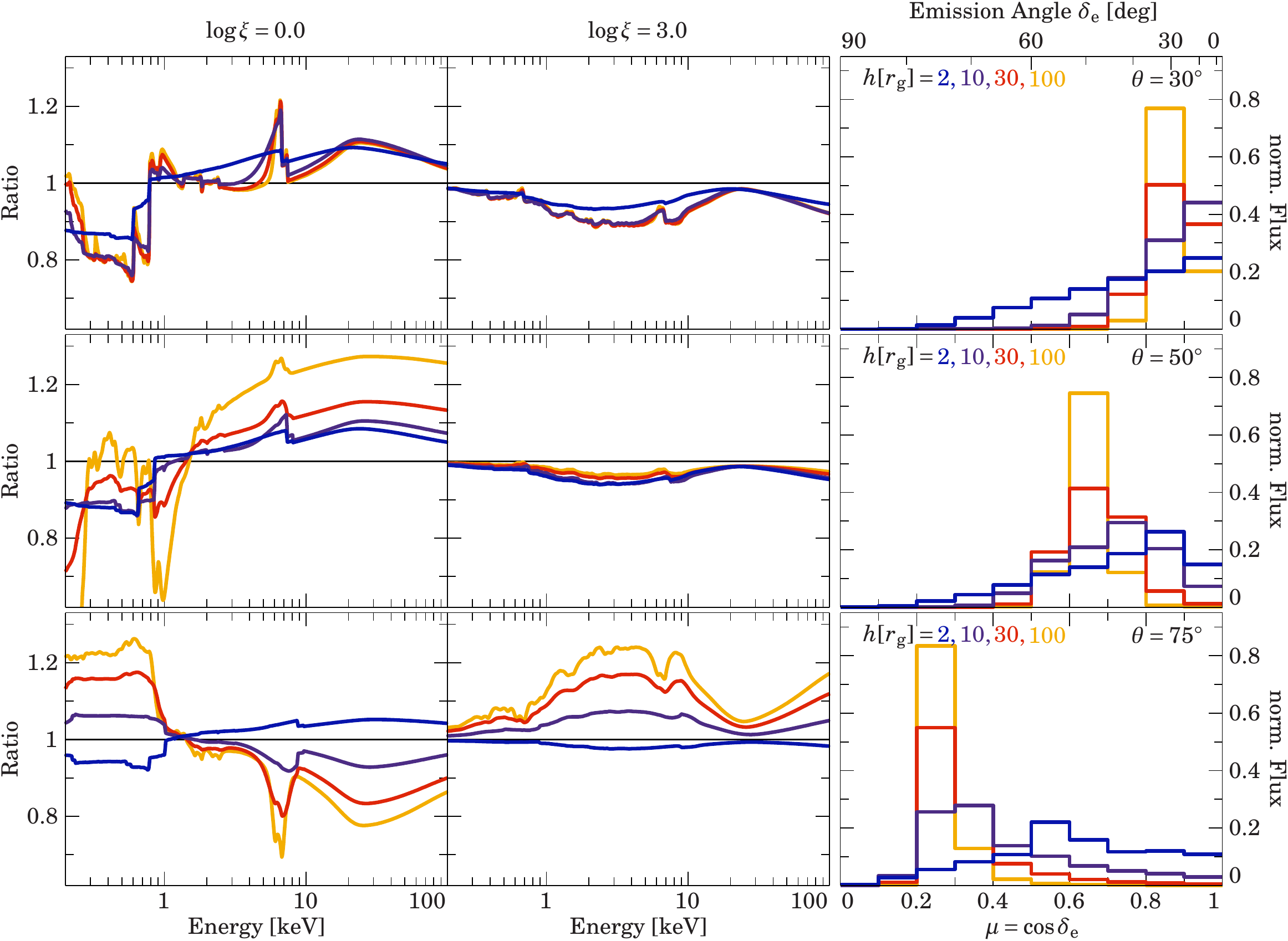}
\caption{The {\tt relxill\_lp} model, which predicts the relativistically
blurred reflected spectrum in a lampost geometry for a combination of
typical parameters. Left and middle panels show the ratio of the new model
to the angle-averaged solution for log~$\xi=0$ and $3$, respectively, while
the right panel shows the distribution of emission angles at which photons
are emitted out of the accretion disk. Top, middle, and bottom panels correspond
to inclination (viewing) angles $\theta=30\degree, 50\degree$ and $75\degree$,
respectively. Different colors indicate the height of the primary source, as
shown in the right panels. 
}
\label{fig:rel_refl_profiles}
\end{figure*}

Figure~\ref{fig:rel_refl_profiles} shows the result of incorporating the
angle-dependent reflection models from {\sc xillver} into lampost geometry,
synthesized with the relativistic blurring code into the end-product {\tt
relxill\_lp} \citep{dau13}. Each row represents a different inclination
angle of the system (as indicated in the left panels). In all of the panels,
each color represent a different height of the illumination source (as
indicated in the middle panels). The left column shows the integrated spectra
reflected from the disk over all considered heights of the source.  Intuitively
and evidently, the closer the proximity of the source to the black hole, the
stronger the relativistic effects that distort the shape of the lines. The
middle column contains the ratio of each spectrum to the solution predicted
using the angle-averaged case. Significant differences can be seen across the
entire energy spectrum. For both very low and very high inclinations, the
discrepancies in the Fe K line region can be as large as 40\%. As expected, the
differences with respect to the angle-averaged solution are smaller for
intermediate inclinations. The right column shows the distribution of emission
angles for each configuration. In other words, these panels show the relative
contribution of photons emitted from different angles to a spectrum
corresponding to an observation at a particular inclination. As expected, the
distribution peaks at the viewing angle, in particular when the source is
located far away from the center. For cases where the height of the
illumination source is low ($h\sim 2-10~\rg$), the distribution can become very
flat, and even more so when the disk is observed at large inclination angles. A
very similar result can be seen for the models with $\log\xi=3$
(Figure~\ref{fig:rel_refl_profiles}, middle column). Notice that when the
distribution of the emission angle becomes very flat, the resultant spectrum is
very similar to the angle-averaged solution (since in this case the observer is
detecting photons coming from all angles with similar contributions, which has
the effect of averaging the signal).

An important caveat to notice is that all the {\sc xillver} models have been
calculated implementing an incident irradiation at $45\degree$ with respect to
the disk normal. As discussed in \cite{gar10} and \cite{dau13}, the incidence
angle affects the specific intensity at the illuminated boundary of the slab,
which influences to certain degree its ionization structure. The assumption is
that the differences introduced by the incidence angle can be mimicked by a
change in the ionization parameter. However, in the context of a lampost
scenario these effects could have a larger impact on the simulated spectrum. A
proper treatment of variable incidence angles requires a much larger
calculation of reflection models which is beyond the scope of the present work,
but it will be implemented in future versions of the model.

\section{Estimating Bias from Angle-Averaged Reflection Results}\label{sec:sim}

As evident from the previous section, significant differences exist when using
the angle-averaged emissivity instead of the proper relativistically lensed
treatment. However, prior to this work, the only available models for data
fitting were angle averaged. Hence, we aim to determine to what extent this
approximate treatment affects the fitted parameters as well as what can be
gained by using the fully relativistic emissivity angle distribution. 

In order to estimate the degree of bias from a typical observation, we produce
a spectrum of an AGN showing relativistic reflection by simulating a 100\,ksec
observation with \textsl{XMM-Newton} assuming an unblurred reflection component
with flux comparable to \mbox{MCG$-$6-30-15} \citep{bre06}. We employ the
lampost version of our new model (\texttt{relxill\_lp}) for the
intrinsic, emitted reflection from which the observation is obtained.
Subsequently, the simulated data is then fit by a commensurate (and common)
angle-averaged model for X-ray reflection, by setting the flag {\tt
angleon=0}\footnote{Note that setting {\tt angleon=0} is equivalent to
implement a relativistically blurred reflection in the common fashion such as
\texttt{relconv\_lp} $\times$ \texttt{xillver}}.

\begin{figure*}
\epsscale{1.0}\plotone{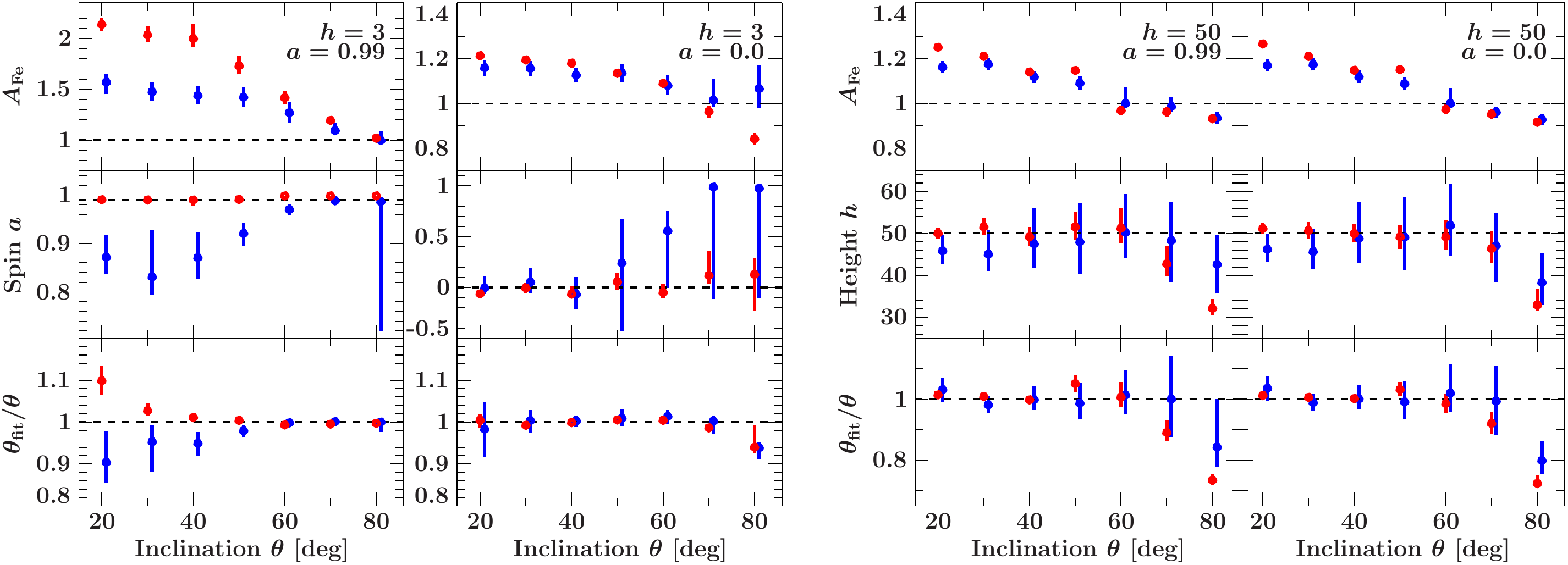}
\caption{ Results from simulating a 100\,ksec \textsl{XMM-Newton} observation
  of an unblurred reflection spectrum assuming the angle-dependent model of 
  \texttt{relxill\_lp}.  We have normalized the flux to be commensurate with 
  \mbox{MCG$-$6-30-15} (i.e., $3.7\cdot10^{-11}\,
  \mathrm{erg}\,\mathrm{s}^{-1}\,\mathrm{cm}^{-2}$). Upon being generated with 
  appropriate noise, the data are
  fitted using the angle averaged model \texttt{relconv\_lp} $\times$
  \texttt{xillver}.   
  The input ionization of the accretion disk is chosen to
  be $\xi=1$ (red, low ionization) or $\xi=10^{3}$ (blue, high
  ionization).  We choose standard initial values of the photon index
  ($\Gamma=2$) and iron abundance ($A_\mathrm{Fe}=1$). All
  combinations are simulated for different input inclinations,
  while we allow all fit parameters (the normalization, $\xi$,
  $a$, $\theta$, $A_\mathrm{Fe}$, and $h$) to vary freely. In the bottom panels, we show
  how the fitted inclination deviates from the input value. \textbf{Left:} Results for a low source height
  ($h=3\,r_\mathrm{g}$), which produces strongly concentrated irradiation
  in the innermost parts of the accretion disk. \textbf{Right:} A
  larger source height ($h=50\,r_\mathrm{g}$), which more strongly illuminates 
  the outer disk. The spin cannot be well
  constrained for such a large height \citep[see][]{dau13}, so we plot
  the deviation in height $h$ rather than spin. }
\label{fig:sim_refl}
\end{figure*}

Figure~\ref{fig:sim_refl} shows the bias estimates obtained from the
simulation. In general, the behavior of the fitting parameters can be divided
into two classes, depending on the irradiation of the accretion disk.  These
classes are delineated by the emissivity profile as (a): showing strong,
centrally-concentrated emissivity which predominantly fluoresces the innermost
accretion disk (a high inner emissivity index), and (b): showing a flattened
emissivity profile with a scaling below $r^{-3}$.  Class (a) is readily
understood as resulting from a low source height, whereas (b) represents a
larger source height which provides a more uniform illumination of the inner
disk.  In the following we will interpret these results in detail. The values
of our bias estimates, for both classes, can be found in Table~\ref{tab:sim}.

In order to study class (a) in which the emissivity is steep
(Figure~\ref{fig:sim_refl}, left panels), we use a source height of $h=3\,\rg$.
In general, it is evident that the iron abundance is most affected. In almost
all the simulations the abundance is over-predicted and generally the effect is
largest for low inclination angles. For a large spin ($a_*=0.99$) and an almost
neutral disk ($\xi=1$), this effect can be more than a factor of two. For high
inclinations this effect is reduced for all parameter configurations, while for
$a_*=0.0$ and $\xi=1$ this turns into the opposite trend.  Especially
interesting is how the angle-averaged treatment influences spin determination.
For the usual AGN observation (large spin, low ionization disk) there seem to
be only very slight deviation from the input parameters. If the ionization is
larger ($\xi=10^{3}$), then stronger biases arise: A large spin ($a_*=0.99$)
will lead to an underestimate when fitting with angle-averaged models for low
inclinations ($\theta<50^\circ$), while for larger inclinations
($\theta>60^\circ$), low spin values ($a_*=0.0$) can be overestimated (but with
large uncertainties). The inclination itself is essentially unaffected by the
angle-averaged treatment. Only in the case of very low inclinations
($\theta<40^\circ$) and high spin, the inferred inclination can be slightly
overestimated (low ionization) or underestimated (high ionization), but the
change is still quite small. 

In the other instance (class b), Figure~\ref{fig:sim_refl} (right panels) shows
the results for a primary irradiating X-ray source placed at $h=50\,\rg$.  For
this particular choice of parameters the spin cannot be very well constrained
\citep[see][]{dau13}, and thus we analyze the effect of the averaged-angle
treatment on the fitted height of the primary source, which is equivalent to
determining the steepness of the emissivity profile.  As in class (a), iron
abundance is generally overestimated. However, the absolute bias is only around
20\% which much lower for (b) compared to the bias obtained from the steep
emissivity of (a).  As before, this effect diminishes for higher inclinations.
For the fitted inclination and the fitted height of the primary source, the
degree of bias is very similar: A significant difference is only obtained for
large inclinations ($\theta>60^\circ$), in the sense that at increasing values
of (input) inclination, the determination of height and the inclination will be
underestimated by angle-averaged models. This effect is again much stronger and
more significant for a low ionization of the accretion disk.

%
%==================================================================================
%
\section{Application to Observational Data}\label{sec:obs}
\begin{figure*}
\epsscale{0.8}\plotone{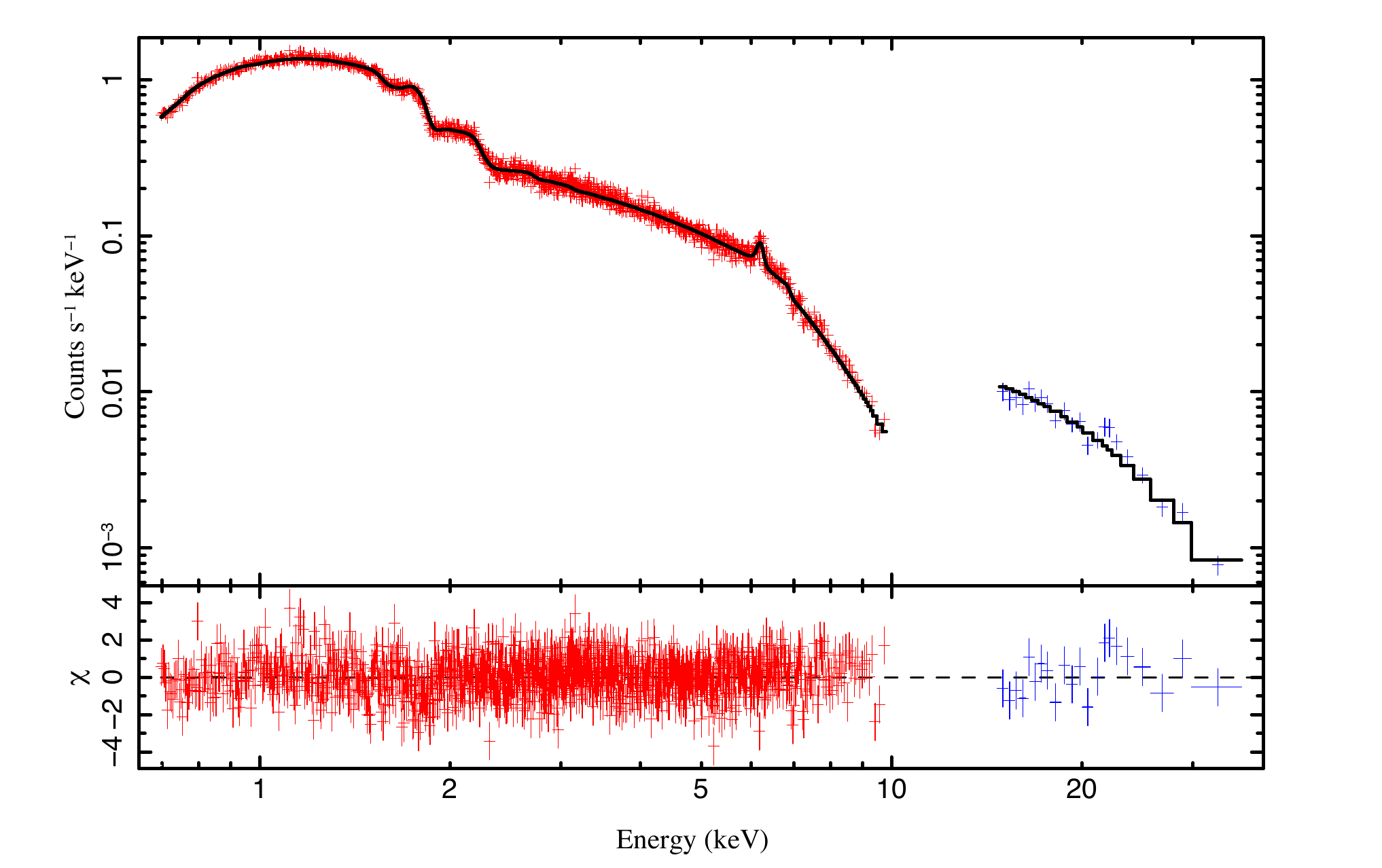}
\caption{{\it Suzaku} spectrum of the Seyfert 1 galaxy Ark~120. Data points
in red show the front-illuminated co added XIS CCD data in the 0.7-10~keV band. 
Blue data points show the PIN data covering the 15-45~keV range. The solid black
line is the best fit using the new {\tt relxill} model (with {\tt angleon=1}).
Residuals are shown in the lower panel.
Error bars are calculated to a 90\% confidence level.
}
\label{ark120}
\end{figure*}

To illustrate how the discussed changes in the model influence the
spectroscopic results of a real dataset, we study the {\it Suzaku} spectrum of
the `bare' Seyfert~1 galaxy Ark\,120. The term `bare' refers to the weak amount
of any intrinsic absorption, which simplifies the spectral analysis
considerably \citep{pat11b}. The $\sim 100$\,ks exposure (ObsID:702014010) was
taken in April 2007 and reduced following the instructions of the {\it Suzaku}
ABC Guide using the newest available calibration. For the spectral analysis,
the two front-illuminated CCDs were co-added and binned to a signal-to-noise of
10. The energies $0.7-10$~keV were noticed for fitting. The PIN data was used
from $15-45$~keV and binned to a signal-to-noise of 5. The back-illuminated XIS
chip (XIS1) was not considered for this analysis. \citet{nar11} and
\citet{wal13} already showed that the spectrum of Ark~120 can be well described
by a simple reflection model, including both a cold and blurred ionized
reflector. Utilizing a similar model set-up to \citet{wal13}, as well as
applying the same restrictions to the parameters, we first describe the
spectrum with the angle-averaged version of our new model {\tt relxill}
(by setting the flag {\tt angleon=0}) for the blurred ionized reflector, and
the cold reflector with a simple {\sc xillver} table (with $\xi$ fixed to 1).
The key parameters of this fit are presented in Table~2. In
particular, we derive a spin of $a_*=0.674_{-0.203}^{+0.118}$ and an
inclination of $i=45.0_{-2.7}^{+5.3}$~deg. To see the influence of the
angle-dependence we then set the flag {\tt angleon=1} for the blurred
reflector, and replace the cold reflector with the angle dependent version of
the {\sc xillver} table. The inclination angles of the two reflectors were
linked for fitting. Similar to the angle-averaged fit the quality is very good
and the spectrum is well described by this model (see Figure~\ref{ark120}).
The angle-dependence improves the spin constraint to
$a_*=0.655_{-0.126}^{+0.122}$ and also improves the constraint on the
inclination slightly ($i=45.3_{-2.4}^{+4.8}$~deg). This is in agreement with
the difference from the angle-averaged model predicted in Section~4.1, which is
of order 10\,\%, and modest compared to the measurement errors.  Error bars
were calculated to a 90\% confidence level.

\begin{figure}
\epsscale{0.8}\plotone{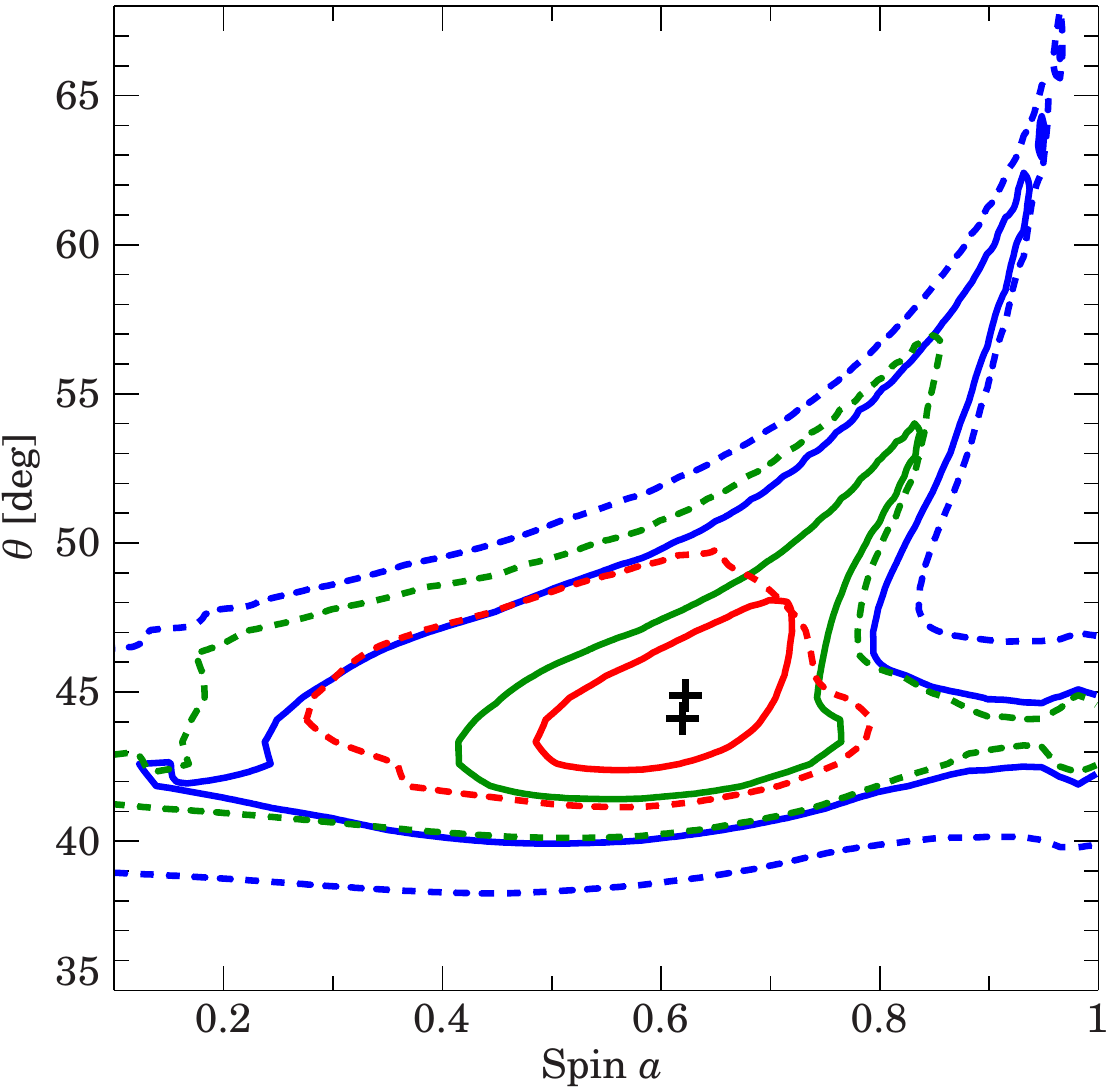}
\caption{
 Contour plots of the inclination and spin parameter $a_*$ for the
 fits to the {\it Suzaku} spectrum of Ark~120. Red, green, and blue 
 solid lines represent the 69\%, 90\%, and 99\% confidence regions for
 the {\tt relxill} fit with {\tt angleon=1}, respectively. The 
 dashed-lines show the same regions (in the same order) for the fit with
 {\tt angleon=0} (angle-averaged solution). The black crosses
 show the position of the minimum $\chi^2$ for each fit, which indicates
 the best-fit parameters. The model parameters are constrained significantly
 better when the new version of the model is implemented.
}
\label{fig:contour}
\end{figure}

To better illustrate the improvement achieved by our new model in constraining
the physical parameters, we have produced the contour plots for the spin 
parameter $a_*$ and the inclination angle. In Figure~\ref{fig:contour} solid lines
show the 69\%, 90\%, and 99\% confidence regions of these two parameters for the 
fit using the angle-resolved ({\tt angleon=1}) option in the new model 
{\tt relxill}. Dashed-lines show the same contours obtained while 
fitting with the angle-averaged solution ({\tt angleon=0}). The crosses show
the position of the minimum value of $\chi^2$ for each case. Although both
fit recover similar best-fit values, the accuracy in the determination of 
both the spin and inclination is significantly higher with the angle-resolved
model. One can see that both parameters are better determined at 90\% level
with the new model than a 69\% level with the old model. With this particular
test case we have shown the capabilities of the new model in the interpretation
of X-ray spectra from accreting sources, as it constitutes a more consistent
and physical description of the reflected spectra from accretion disks around
black holes.

Evidently, this is just one particular example of the potential of the
new model in fitting X-ray observations significantly improved: the
errors in both spin and inclination are better

\section{Summary and Conclusions}\label{sec:con}

The solution of the angular distribution of X-ray reflection from an accretion
disk has been calculated with our reflection code {\sc xillver}. The reflected
spectrum at any given angle shows clear discrepancies of up to $\sim 25 \%$
when compared to the angle-averaged solution. This is observed to occur for all
angles, and at any ionization parameter. In particular, scattered continuum
photons and line photons respond differently to the angular effects, since the
optical depth can be significantly different at different energies.  Notably,
the Fe K emission line complex becomes weaker for high viewing angles, which
agrees with earlier claims of the decrease of the line equivalent width when
observed at grazing angles \citep[e.g.][]{bas78,mat92,mar00,rey04}.  The
angular effects are introduced in the reprocessed radiation mainly by
changing the effective optical depth $\tau_\mathrm{eff} = \tau/\mu$, where
$\mu$ is the cosine of the viewing angle. The angular effects are much more
extreme at energies where the photoelectric absorption dominates over the
Compton scattering. At high energies ($\gtrsim 20$~keV), Compton scattering is
the only source of opacity, and thus the angular effects are independent of the
ionization of the gas. An important consequence of these effects is that they
change the slope of the reflected spectrum as the angular-dependency acts 
differently at different energies. 

The new reflection models predict a limb-brightening law, i.e., the reflected
flux integrated over all energies becomes larger at high inclination angles,
which agrees well with hydrostatic calculations of \cite{roz08} and \cite{roz11}. 
However, its
functional form does not agree with the commonly used law $I \propto
\ln(1+1/\mu)$ \citep{haa93}, instead, our calculations show a profile less
steep. When the atmosphere is highly ionized (log~$\xi > 3$), the profile
becomes even flatter, approaching to the isotropic case. By carrying out
calculations of a pure scattering atmosphere, we have shown that this is
expected as Compton scattering is essentially isotropic for the reflected
spectra of irradiated plane-parallel slabs.  In the context of {\it unblurred
reflection}, we expect that these models will be most relevant in analyzing
data from astrophysical systems at either very low or very high inclination
angles. A new {\sc xillver} table which includes the inclination as a model
parameter is provided in the proper
format\footnote{\url{http://hea-www.cfa.harvard.edu/~javier/xillver/}} for its
implementation via the {\tt atable} model in {\sc xspec}.

The table of reflection spectra mentioned above is intended to be used to model
an unblurred reflection component, as is the case of the distant cold
reflection observed in many AGN. However, it is important to emphasize that when
relativistic effects need to be taken into account, a convolution with blurring
models in the usual fashion (e.g. {\tt relconv $\times$ xillver}) is in fact
incorrect, as the proper link of the two models needs to be done
intrinsically at each radial zone. Instead, we advise observers to implement
the newly developed model {\tt
relxill}\footnote{\url{http://www.sternwarte.uni-erlangen.de/research/relxill/}},
which properly integrates the two models. An additional version of this model
({\tt relxill\_lp}), in which the illumination is prescribed under a {\it
lampost} scenario, is also provided. The relativistic effects introduce
additional directionality dependences in the final reflected spectrum. In
general, deviations from the angle-averaged solution can be as large as $20\%$,
and like in the unblurred reflection case, lines and continuum respond in a
different way. The main effect is that, at any given inclination, a distant
observer will not only see those photons that are emitted into its line of
sight, but also those emitted at different directions, but for which the light
bending effects are enough to modify their path so that they eventually point
toward the observer (for instance, at grazing angles, an observer can even see
photons emitted behind the black hole, due to the extreme space-time
curvature). 

Models calculated in a lampost geometry, where the illumination of
the disk is mostly prescribed by the height of the source over the
accretion disk, also differ significantly from their angle-averaged
counterparts. We found that for either very low or very high inclinations,
the discrepancies in the Fe K line band can be as large as $40\%$.
It is worthwhile to mention that all these relative differences
are for the reflection spectrum only. In real observations, the
observed spectrum is a combination of the reflected and the direct 
component which illuminates the disk. In many cases the reflection
spectrum may only contribute $10-20\%$ of the $2-10$~keV flux of the
direct continuum, which dilutes the reflection features and consequently
the deviations seen between the new and the old versions of the models.
In cases where the height of the illumination source is low ($h\sim 2-10~r_g$),
the distribution of emission angles that are actually observed at a
given inclination becomes very flat. This degree of angular mixing is most
pronounced when the disk is observed at high inclination angles. Accordingly,
when relativistic effects are extreme, an observer is detecting photons from
almost all angles with similar contribution, which essentially acts as an
averaging of the reflected spectra over all angles.

We have also carried out a detailed error analysis of the physical
parameters derived from fitting these models to simulated data. 
Our main goal is to show an estimate of the bias introduced
by the implementation of the approximate (and commonly used) angle-averaged
solution for the reflected spectra. By looking at two distinctive 
cases, i.e., steep and flat irradiation of the disk, we found that
the Fe abundance tends to be over-predicted in almost all cases, but
in particular for low inclinations. For a large spin and low ionization,
the differences can be as large as a factor of two. This could explain
the high abundances usually observed in the spectra of many AGN
\citep[see for example,][]{rey97}. Angular effects also influence
the spin determination, although the effect is not dramatic for typical
parameters. In particular, for large spin and low ionization
the spin recovered from the fits is underestimated at low inclinations
($\theta < 50\degree$), while for larger inclinations originally low
spins ($a_*=0.0$) are overestimated. The bias on the inclination 
and height of the primary source is only significant in the case
of low spin and large inclination angles, for which both quantities
can be underestimated. Oddly, the determination of the inclination angle
is mostly unaffected when the spin is large.

We have also included an illustrative example of the application 
of the new model to real observational data. For this, we analyzed
the {\it Suzaku} spectrum of the Seyfert 1 galaxy Ark~120. The same
fit is performed using the angle-averaged solution and the complete
angle-dependent model. The differences in the recovered parameters
are modest. Nevertheless, the differences in the constrain of the 
physical parameters is quite significant. The accuracy of the new 
model in the determination of both the inclination and the spin parameter
$a_*$ at a 90\% confidence level is superior than the old, angle-averaged
model at 69\% confidence level. This test case demonstrates the capabilities 
of the new model, which is more consistent and physical than previous versions.
It also illustrates its relevance in the interpretation of observational 
data from accreting sources, and we expect it will have an even greater impact
on analyzing higher quality data from the new generation of observatories such
as {\it NuSTAR} and {\it Astro-H}.

The models presented here are a new step toward a more physical and
self-consistent representation of the X-ray reflection from accretion disks.
However, additional sources of systematics that could affect the retrieval 
of physical parameters (e.g. black hole spin) need still to be further 
investigated. Although recent studies suggest that for some sources the emission region is compact
\citep[e.g.][]{dau13}, the exact geometry of the primary source of X-rays,
whether it is a corona or the base of jet, is not very well understood. The
assumptions made on this respect affect the way X-rays illuminate the the
accretion disk. By providing a lamppost version of our reflection model we are
taking the first steps to unveil the geometry of the illuminating source.
However, further studies \citep[e.g., timing measurements;
see][]{zog10,kar13a,kar13b}, are required to accurately determine the
illumination profile and the physical origin of the primary source itself. The
vertical ionization structure of the disk is treated in detail with the models
presented here, however, the same ionization parameter is assumed for all
radii. The radial profile of the ionization depends on both the illumination
and the gas density profiles. This will be subject of future studies. Other
sources of systematics are not directly related to the reflection spectrum. For
instance, the presence of a warm absorber in several AGN is known to modify the
prediction of the fits depending on how this component is being treated
\citep[e.g., MCG-6-30-15, ][]{bre06}. However, observations carried out with
the new observatory {\it NuSTAR} in combination with {\it XMM-Newton} show that
we are currently on the edge of being able to distinguish between blurred
reflection and absorption \citep[with NGC~1365 being the best example
so far,][]{ris13}.  The improved physical description of the illuminated gas is
expected to play a major role to disentangle these components by reducing the
uncertainties in the X-ray reflection spectrum.
%
%
%
%==================================================================================
%
%
\acknowledgments

The calculations presented here were performed in the Odyssey cluster
of the Research Computing Facilities of the Harvard University.
JG and JEM acknowledge the support of NASA grant NNX11AD08G.
JFS was supported by NASA Hubble Fellowship grant HST-HF-51315.01.
CSR thanks support from NASA under grant NNX10AE41G.
%
%
%==============================================================================
%
\bibliographystyle{apj}
\bibliography{my-references}
%
%==================================================================================
%
\begin{table*}
  \caption{Fractional bias of the parameters spin, iron abundance and 
    height when using the angle-averaged approximation for the reflected 
    radiation. The difference between the original value and the one recovered
    after the fit is shown for each parameter. All uncertainties are
    given at 90\% confidence.}
  \label{tab:sim}
  \scriptsize
\begin{center}
\begin{tabular}{lcc|rrrr|rrrrr|r}
\tableline\tableline 
 &  &  & \multicolumn{4}{c}{ --- $\log\xi = 0.0$ --- }& \multicolumn{4}{c}{ --- $\log\xi = 3.0$ --- }\\
 & $a_*$ & $h$ & $\theta=20^\circ$& $\theta=40^\circ$& $\theta=60^\circ$& $\theta=80^\circ$& $\theta=20^\circ$& $\theta=40^\circ$& $\theta=60^\circ$& $\theta=80^\circ$\\ \tableline
$A_\mathrm{Fe}$ & 0.99 & 3 & $1.14\pm0.07$& $1.00^{+0.15}_{-0.09}$& $0.42\pm0.07$& $0.02\pm0.03$& $0.57^{+0.09}_{-0.12}$& $0.44^{+0.10}_{-0.09}$& $0.27\pm0.11$& $-0.00^{+0.10}_{-0.02}$\\
$a_*$           & 0.99 & 3 & $0.00\pm0.01$& $-0.00^{+0.01}_{-0.02}$& $0.01^{+0.00}_{-0.01}$& $0.01^{+0.00}_{-0.01}$& $-0.12^{+0.05}_{-0.04}$& $-0.12^{+0.06}_{-0.05}$& $-0.02\pm0.02$& $-0.00^{+0.01}_{-0.27}$\\
$\theta$        & 0.99 & 3 & $1.97^{+0.69}_{-0.67}$& $0.44^{+0.31}_{-0.36}$& $-0.39^{+0.18}_{-0.20}$& $-0.23^{+0.20}_{-0.16}$& $-1.92^{+1.51}_{-1.00}$& $-2.03^{+1.08}_{-1.16}$& $-0.09^{+0.43}_{-0.48}$& $0.00^{+0.00}_{-1.86}$ \\
\tableline
$A_\mathrm{Fe}$ & 0.00 & 3 & $0.21\pm0.02$& $0.18^{+0.02}_{-0.03}$& $0.09\pm0.02$& $-0.16\pm0.03$& $0.16\pm0.04$& $0.13\pm0.04$& $0.08\pm0.05$& $0.07^{+0.11}_{-0.09}$\\
$a_*$           & 0.00 & 3 & $-0.06^{+0.04}_{-0.03}$& $-0.06^{+0.08}_{-0.05}$& $-0.05^{+0.09}_{-0.06}$& $0.13^{+0.17}_{-0.36}$& $-0.00^{+0.12}_{-0.07}$& $-0.07^{+0.17}_{-0.15}$& $0.56^{+0.20}_{-0.57}$& $0.97^{+0.03}_{-1.09}$\\
$\theta$        & 0.00 & 3 & $0.09^{+0.31}_{-0.39}$& $-0.04^{+0.14}_{-0.20}$& $0.27^{+0.20}_{-0.19}$& $-4.79^{+4.16}_{-1.13}$& $-0.34^{+1.30}_{-1.34}$& $0.11^{+0.49}_{-0.59}$& $0.81^{+0.91}_{-1.06}$& $-4.91^{+1.03}_{-2.20}$ \\
\tableline
$A_\mathrm{Fe}$ & 0.99 & 50 & $0.25\pm0.02$& $0.14\pm0.02$& $-0.03\pm0.03$& $-0.07^{+0.03}_{-0.02}$& $0.16\pm0.03$& $0.12\pm0.03$& $-0.00^{+0.08}_{-0.01}$& $-0.06\pm0.03$\\
$a_*$           & 0.99 & 50 & $-1.99^{+0.37}_{-0.01}$& $-1.99^{+1.01}_{-0.00}$& $0.01^{+0.00}_{-1.76}$& $0.01^{+0.00}_{-0.03}$& $-1.99^{+2.00}_{-0.01}$& $-1.98^{+1.99}_{-0.01}$& $0.01^{+0.00}_{-2.00}$& $0.00^{+0.01}_{-2.00}$\\
$\theta$        & 0.99 & 50 & $0.29\pm0.20$& $-0.07^{+0.44}_{-0.42}$& $0.43^{+2.95}_{-1.99}$& $-21.22^{+1.60}_{-1.23}$& $0.62^{+0.79}_{-0.82}$& $-0.06^{+1.81}_{-1.33}$& $0.80^{+4.90}_{-3.74}$& $-12.55^{+12.55}_{-5.09}$\\
\tableline
$A_\mathrm{Fe}$ & 0.00 & 50 & $0.27\pm0.02$& $0.15\pm0.02$& $-0.03\pm0.03$& $-0.08^{+0.03}_{-0.02}$& $0.17\pm0.03$& $0.12\pm0.03$& $0.00^{+0.07}_{-0.02}$& $-0.07\pm0.03$\\
$a_*$           & 0.00 & 50 & $-1.00^{+0.49}_{-0.00}$& $-1.00^{+0.90}_{-0.00}$& $1.00^{+0.01}_{-1.55}$& $1.00^{+0.00}_{-0.04}$& $-1.00^{+2.00}_{-0.00}$& $-1.00^{+2.00}_{-0.01}$& $0.77^{+0.23}_{-1.78}$& $0.99^{+0.01}_{-1.99}$\\
$\theta$        & 0.00 & 50 & $0.25\pm0.20$& $0.10^{+0.46}_{-0.41}$& $-0.86^{+1.95}_{-1.78}$& $-22.10^{+2.19}_{-0.86}$& $0.72^{+0.80}_{-0.82}$& $0.03^{+1.84}_{-1.35}$& $1.20^{+5.80}_{-3.65}$& $-16.09^{+5.17}_{-3.49}$\\\tableline 
\end{tabular}
\end{center}
\end{table*}
%
%==================================================================================
%
\begin{deluxetable}{lccc}
\label{tfit}
\tabletypesize{\scriptsize}
\tablecaption{Best-fit parameters for Ark~120 data}
\tablewidth{0pt}
\tablehead{
\colhead{Component} & \colhead{Parameter} & \colhead{Value} & \colhead{Value} \\
}
\startdata
{\tt relxill} & $q_{in}=q_{out}$ & $4.4_{-1.2}^{+2.0}$       & $4.8_{-1.1}^{+1.8}$ \\
{\tt relxill} & $a$              & $0.674_{-0.203}^{+0.118}$ & $0.655_{-0.126}^{+0.122}$ \\
{\tt relxill} & $i$ (deg)        & $45.0_{-2.7}^{+5.3}$      & $45.3_{-2.4}^{+4.8}$ \\
{\tt relxill} & $R_{in}$ (ISCO)  & $1$                       & $1$ \\
{\tt relxill} & $R_{out}$ (ISCO) & $400$                     & $400$ \\
{\tt relxill} & $z$              & $0.0327$                  & $0.0327$ \\
{\tt relxill} & $\Gamma$         & $2.17_{-0.01}^{+0.02}$    & $2.17_{-0.01}^{+0.02}$ \\
{\tt relxill} & log~$\xi$        & $1.03_{-0.26}^{+0.16}$    & $0.84_{-0.10}^{+0.24}$ \\
{\tt relxill} & $A_{Fe}$         & $1.54_{-0.45}^{+0.44}$    & $1.78_{-0.40}^{+0.39}$ \\
{\tt relxill} & {\tt angleon}  & 0                         & 1 \\
{\tt relxill} & $N (10^{-4})$    & $1.81_{-0.29}^{+0.32}$    & $2.07_{-0.2}$ \\ 
{\tt xillver}       & $N (10^{-4})$    & $1.55_{-0.40}^{+0.42}$    & $0.14_{-0.04}^{+0.07}$ \\
\enddata
\end{deluxetable}
\end{document}